\documentclass[preprint,12pt]{elsarticle}
\journal{International Journal of Engineering Science}
\usepackage{graphics,epsfig}
\usepackage{graphicx}
\usepackage{float}
\usepackage{amssymb,amstext,amsmath}
\usepackage{mathtools}
\usepackage{booktabs}
\usepackage{physics}
\usepackage{comment}
\usepackage{appendix}
\usepackage{listings}
\begin{document}
\begin{frontmatter}
\title{Asymptotically exact theory of functionally graded elastic beams}
\author{K. C. Le$^{a,b}$\footnote{Email: lekhanhchau@tdtu.edu.vn}, T. M. Tran$^c$}
\address{$^a$Division of Computational Mechanics, Institute for Advanced Study in Technology, Ton Duc Thang University, Ho Chi Minh City, Vietnam\\
$^b$Faculty of Civil Engineering, Ton Duc Thang University, Ho Chi Minh City, Vietnam
\\
$^c$Faculty of Mechanical Engineering, Vietnamese German University,
Binh Duong, Vietnam}
\begin{abstract} 
We construct a one-dimensional first-order theory for functionally graded elastic beams using the variational-asymptotic method. This approach ensures an asymptotically exact one-dimensional equations, allowing for the precise determination of effective stiffnesses in extension, bending, and torsion via numerical solutions of the dual variational problems on the cross-section. Our theory distinguishes itself by offering a rigorous error estimation based on the Prager-Synge identity, which highlights the limits of accuracy and applicability of the derived one-dimensional model for beams with continuously varying elastic moduli across the cross section.
\end{abstract}


\end{frontmatter}

\section{Introduction}
Functionally graded (FG) materials \cite{niino1990recent,koizumi1992recent}, with their spatially varying properties, offer tremendous potential for designing advanced structures with tailored performance. While FG materials have seen widespread use in various applications \cite{saleh202030}, accurately modeling their behavior, particularly in beam structures, remains a challenge.  The exact treatment of FG structures within three-dimensional (3-D) elasticity is only possible in a few exceptional cases due to the complicated boundary value problems \cite{mian1998exact,horgan1999pressurized,sankar2001elasticity,pan2003exact,kashtalyan2004three,zenkour2007benchmark,chu2015two}.  Classical one-dimensional (1-D) beam theories \cite{sankar2001elasticity,aydogdu2007free,csimcsek2009free} and higher-order 1-D beam theories \cite{li2008unified,thai2012bending,bourada2015new,reddy2021theories} often rely on simplifying assumptions that don't fully capture the complexity of FG materials, especially the variation of their elastic properties. This can lead to inaccuracies in predicting their structural response, particularly when the material properties including Poisson's ratio vary significantly across the beam's cross-section.

This paper presents a rigorous approach to develop an asymptotically exact one-dimensional theory for FG beams using the variational-asymptotic method (VAM), a powerful technique for systematically reducing the dimension of structural models while retaining essential accuracy \cite{berdichevsky1979variational}. Berdichevsky \cite{berdichevskii1981energy} was the first to apply VAM to derive the 1-D first order beam theory from 3-D elasticity theory. His asymptotic analysis reveals that the static (or dynamic) three-dimensional problem of an inhomogeneous beam can be decoupled into two lower-dimensional problems: (i) a two-dimensional cross-sectional analysis and (ii) a one-dimensional variational problem. The solution of the cross-sectional problem is crucial as it determines the energy functional for the 1-D variational problem. However, Berdichevsky's work only addressed the cross-sectional problem for materials with a constant Poisson's ratio, limiting the general applicability of the dimensional reduction. Our paper aims to overcome this limitation by providing a comprehensive analysis of the cross-sectional problem for FG beams with arbitrary material properties. Our approach employs a finite element implementation to solve the dual variational problems on the cross-section, ensuring convergence to the true minimizers. Furthermore, we establish clear relationships between the 3-D stress state and the 1-D beam characteristics, enabling accurate prediction of the beam's behavior. Finally, we provide an error estimation of the 1-D theory, establishing its accuracy and reliability. 

The paper is structured as follows. After this short introduction, the variational formulation of the problem is given in Section 2. Section 3 is devoted to the asymptotic analysis of the 3-D energy functional of FG beams, which leads to the cross-sectional problems. In Section 4, the cross-sectional problems and their dual formulations are solved using the finite element method. Section 5 presents the 1-D theory and the relationship between its solution and the 3-D stress and strain states. Section 6 formulates and proves the error estimation based on the Prager-Synge identity. Finally, Section 7 concludes the paper.

\section{Variational formulation for functionally graded beams}
\begin{figure}[htb]
	\centering
	\includegraphics[width=12cm]{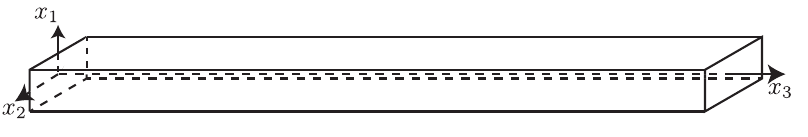}
	\caption{A beam}
	\label{fig:1}
\end{figure}
Let us consider a functionally graded (FG) beam occupying the region $\mathcal{B}=\mathcal{A}\times (0,L)$ in its undeformed state, where $\mathcal{A}$ represents the cross-sectional domain in the $(x_1,x_2)$-plane (see Fig.~\ref{fig:1} presenting an example of a beam with a rectangular cross section). Let $x_3\equiv x$ be the coordinate along the beam axis. In addition, we choose the origin of the $(x_1,x_2)$-coordinates so that it matches the centroid of $\mathcal{A}$. We assume the beam's material properties, characterized by Lamé's moduli $\lambda$ and $\mu$, vary continuously across the cross-section. The beam is clamped at $x=0$, and subjected to a specified traction $\mathbf{t}(x_1,x_2)$ at $x=L$.

According to Gibbs' variational principle \cite{berdichevsky2009variational}, the true displacement field $\check{\mathbf{u}}(\mathbf{x})$ of the beam in equilibrium minimizes the energy functional
\begin{equation}
\label{eq:1}
I[\mathbf{u}(\mathbf{x})]=\int_{\mathcal{B}}W(x_1,x_2,\vb*{\varepsilon})\dd[3]{x}-\int_{\mathcal{A}} \mathbf{t}(x_1,x_2)\vdot \mathbf{u}(x_1,x_2,L) \dd[2]{x}
\end{equation}
among all kinematically admissible displacements $\mathbf{u}(\mathbf{x})$ that vanish at $x=0$. Here, $\dd[3]{x}=\dd x_1\dd x_2\dd x$ is the volume element, $\dd[2]{x}=\dd x_1\dd x_2$ the area element, $W(x_1,x_2,\vb*{\varepsilon})$ is the stored energy density given by
\begin{equation*}
W(x_1,x_2,\vb*{\varepsilon})=\frac{1}{2}\lambda(x_1,x_2) (\tr \vb*{\varepsilon})^2+\mu(x_1,x_2) \vb*{\varepsilon} \mathbf{:}\vb*{\varepsilon},
\end{equation*}
where $\vb*{\varepsilon} = \frac{1}{2}(\grad \mathbf{u}+(\grad \mathbf{u})^T)$ is the strain tensor.

We aim to simplify the analysis of a thin FG elastic beam by replacing the full three-dimensional energy functional \eqref{eq:1} with an approximate one-dimensional functional (dimension reduction). This dimension reduction exploits the geometry of the slender beam, where the characteristic size of the cross-section, $h$, is significantly smaller than the length, $L$. By employing a variational-asymptotic method, we will derive a one-dimensional functional that accurately captures the beam's behavior while neglecting terms of order $h/L$ compared to unity. This constitutes a first-order, or ``classical'', approximation, formally corresponding to the limit $h\to 0$.  

To facilitate the variational-asymptotic analysis, we adopt index notation: Greek indices (ranging from 1 to 2) represent components in the transverse $(x_1,x_2)$-plane, while Latin indices denote components in the full 3-D Cartesian coordinates. Summation over repeated indices is implied. For brevity, we omit the index 3 for the longitudinal coordinate $x_3$, the corresponding displacement $u_3$, and the traction $t_3$. 
To maintain a transverse domain as $h\to 0$, we introduce dimensionless coordinates:
\begin{equation*}
y_\alpha =\frac{x_\alpha }{h}, \quad (y_1,y_2) \in \bar{\mathcal{A}},
\end{equation*}
where $\bar{\mathcal{A}}$ represents the fixed normalized cross-sectional domain. This transforms the beam's energy functional to:
\begin{equation}\label{eq:2}
I[u_i(\vb{y},x)]=\int_{0}^{L}\int_{\bar{\mathcal{A}}}h^2W(\vb{y} ,\varepsilon_{ij}) \dd[2]{y} \dd{x} -\int_{\bar{\mathcal{A}}} h^2t_i(\vb{y}) u_i (\vb{y},L) \dd[2]{y}.
\end{equation}
Here, the dimensionless transverse coordinates $y_\alpha$ act as ``fast'' variables compared to the ``slow'' longitudinal coordinate $x$.  We use the vector notation $\vb{y}$ in function arguments to avoid cumbersome index repetition. By expressing the displacements $u_i$ as functions of both $\vb{y}$ and x, we explicitly separate the fast and slow scales. The thickness $h$ now appears explicitly in the strain tensor components:
\begin{equation}\label{eq:3} 
\varepsilon _{\alpha \beta }=\frac{1}{h}u_{(\alpha ;\beta )}, \quad
2\varepsilon _{\alpha 3}= 
\frac{1}{h}u_{;\alpha }+u_{\alpha,x},
\quad
\varepsilon_{33}=u_{,x}.
\end{equation}
Here, a semicolon preceding a Greek index denotes differentiation with respect to the corresponding fast coordinate, and parentheses enclosing a pair of indices indicate symmetrization.

\section{Dimension reduction}

To facilitate the asymptotic analysis of the energy functional as $h\to 0$, we first recast the stored energy density into a more convenient form, following the approach developed by Berdichevsky \cite{berdichevsky2009variational} and Le \cite{le1999vibrations}. The asymptotically dominant terms in $W(\vb{y}, \varepsilon_{ij})$ involve the derivatives $w_{(\alpha ;\beta )}/h$ within $\varepsilon _{\alpha \beta}$ and $w_{;\alpha }/h$ within $\varepsilon _{\alpha 3}$. To emphasize these components, we decompose the stored energy density into three quadratic forms:
\begin{equation*}
W_\parallel =\min_{\varepsilon _{\alpha \beta},
\varepsilon _{\alpha 3}}W, 
\quad
W_\angle =\min_{\varepsilon _{\alpha \beta}} (W-W_\parallel ), 
\quad
W_\perp =W-W_\parallel -W_\angle .
\end{equation*}
We refer to $W_\parallel $, $W_\angle $, and $W_\perp $ as the ``longitudinal'', ``shear'', and ``transverse'' energy densities, respectively. $W_\parallel $ depends solely on $\varepsilon _{33}$ and corresponds to the energy when the stresses $\sigma _{\alpha \beta }$ and $\sigma _{\alpha 3}$  vanish. $W_\angle $ depends only on $\varepsilon _{\alpha 3}$, while $W_\perp $ accounts for the remaining energy contributions.

From these definitions, we obtain:
\begin{equation*}
\begin{split}
W_\parallel &=\frac{1}{2}E(\varepsilon _{33})^2, \quad 
E=\frac{\mu (3\lambda +2\mu )}{\lambda +\mu },
\\ 
W_\angle &=2\mu \varepsilon 
_{\alpha 3}\varepsilon _{\alpha 3}, 
\\
W_\perp &=\frac{1}{2}\lambda (\varepsilon _{\alpha \alpha}
+2\nu \varepsilon _{33})^2+ \mu (\varepsilon _{\alpha \beta }
+\nu \delta _{\alpha \beta }\varepsilon _{33}) (\varepsilon _{\alpha \beta }
+\nu \delta _{\alpha \beta }\varepsilon _{33}),\quad \nu=\frac{\lambda}{2(\lambda+\mu)},
\end{split}
\end{equation*}
where $E$ is Young's modulus and $\nu $ is Poisson's ratio. Recall that $E$, $\nu$, and Lamé's moduli $\lambda$ and $\mu$ are functions of the fast coordinates $\vb{y}$ for FG beams. The energy densities $W_\angle $ and $W_\perp $ can also be expressed as follows:
\begin{equation}\label{eq:4}
\begin{split}
W_\angle &= \sigma_{\alpha 3}\varepsilon_{\alpha 3},  \\
W_\perp &= \frac{1}{2} (\sigma_{\alpha \beta} \varepsilon_{\alpha \beta}+\sigma_{33}\varepsilon_{33}-E(\varepsilon _{33})^2). 
\end{split}
\end{equation}

A formal variational-asymptotic analysis as $h\to 0$ would involve determining a set of functions $\mathcal{N}$ (representing independent degrees of freedom) according to the general scheme outlined in \cite{le1999vibrations}. This procedure would reveal that, to leading order, the displacements $u_i$ are independent of the fast coordinates $\vb{y}$, i.e., $u_i=v_i(x)$. Subsequent iterations would introduce a linear dependence on $\vb{y}$ and a twist angle $\varphi (x)$, ultimately leading to $\mathcal{N}=\{ v_i(x),\varphi(x)\}$. To expedite the derivation, we bypass these standard steps and directly introduce the following Ansatz:
\begin{equation}\label{eq:5}
\begin{split}
&u_\alpha (\vb{y},x)=v_\alpha (x)-he_{\alpha \beta}\varphi (x)y_\beta+hw _\alpha (\vb{y},x),
\\
&u(\vb{y},x)=v(x)-h v_{\alpha ,x} (x) y_\alpha +hw(\vb{y},x),
\end{split} 
\end{equation}
where $e_{\alpha \beta }$ is the two-dimensional permutation symbol. Without loss of generality, we impose the following constraints on the functions $w_\alpha $ and $w$:
\begin{equation}\label{eq:6}
\begin{split}
\langle w_\alpha \rangle =0,\quad  e_{\alpha \beta 
} \langle w_{\alpha ;\beta }\rangle =0, \\
\quad \langle w \rangle =0, \quad  \langle . \rangle \equiv \int_{\bar{\mathcal{A}}} . \dd[2]{y}.
\end{split}
\end{equation}
These constraints ensure that $v_\alpha (x)$ and $v(x)$ represent the mean displacements of the beam's cross-section, while $\varphi (x)$ corresponds to its mean rotation about the beam axis. Equations \eqref{eq:5} and \eqref{eq:6} establish a one-to-one correspondence between the original displacement field $\{ u_\alpha ,u\}$ and the new set of functions $\{v_\alpha ,v, \varphi, w_\alpha ,w\}$.

Invoking the Saint-Venant principle \cite{berdichevskii1974proof,gregory1984decaying}, we decompose the beam's 3-D domain into an inner region and boundary layers of width $O(h)$ near the edges. Within these boundary layers, the stress and strain fields are inherently three-dimensional. Consequently, the energy functional \eqref{eq:2} can be separated into an inner functional, amenable to our asymptotic analysis, and a boundary layer functional. To leading order, we can neglect the boundary layer functional when determining $ w_\alpha $ and $w$. Thus, the dimension reduction problem reduces to finding the minimizers $\check{w}_\alpha$ and $\check{w}$ of the inner functional, which, as $h\to 0$, coincides with functional \eqref{eq:2} without the traction term.
  
We substitute \eqref{eq:5} into the inner functional (i.e., \eqref{eq:2} without the traction term) and retain only the asymptotically dominant terms involving $w_\alpha$ and $w$. The estimations based on Eqs.~\eqref{eq:3} and \eqref{eq:5} lead to the following asymptotic formulas for the strain components:
\begin{equation}
\varepsilon _{\alpha \beta }=w_{(\alpha ;\beta )},\quad 2\varepsilon _{\alpha 3}=
w_{;\alpha }-he_{\alpha \beta}\Omega y_\beta,\quad \varepsilon _{33}=\gamma+h\Omega_\alpha y_\alpha,
\label{eq:7}
\end{equation}
where we have introduced the elongation, bending, and twist measures:
\begin{equation}\label{eq:8}
\gamma=v_{,x},\quad \Omega_\alpha =-v_{\alpha,xx},\quad \Omega=\varphi_{,x}.
\end{equation}
Since the derivatives of $w_\alpha$ and $w$ with respect to $x$ do not appear in the leading-order energy functional, we can treat $x$ as a parameter and focus on minimizing the following decoupled 2-D functionals for each fixed $x$:
\begin{multline}\label{eq:9}
I_\perp [w_\alpha (\vb{y})]=h^2 \langle \frac{1}{2}\lambda(\vb{y}) [w_{\alpha ;\alpha } 
+2\nu(\vb{y}) (\gamma +h\Omega _\sigma y_\sigma )]^2 
\\
+\mu(\vb{y}) [w_{(\alpha ;\beta )} +\nu(\vb{y}) \delta _{\alpha \beta }(\gamma +h\Omega _\gamma y_\gamma )][w_{(\alpha ;\beta )} +\nu(\vb{y}) \delta _{\alpha \beta }(\gamma +h\Omega _\delta y_\delta )]\rangle ,
\end{multline}
\begin{equation}\label{eq:10} 
I_\angle [w(\vb{y})] =h^2 \langle \frac{1}{2}\mu(\vb{y}) (w_{;\alpha }-h\Omega e_{\alpha 
\beta}y_\beta)(w_{;\alpha }-h\Omega e_{\alpha \gamma}y_\gamma)\rangle ,
\end{equation}
where $w_\alpha$ and $w$ satisfy the constraints \eqref{eq:6} and their dependence on $x$ is suppressed for brevity. We refer to the minimization of \eqref{eq:9}, involving only $w_\alpha(\vb{y})$, as the ``plane strain problem'' and the minimization of \eqref{eq:10}, involving only $w(\vb{y})$, as the ``anti-plane'' problem. This decoupling arises from the assumed plane of elastic symmetry perpendicular to the beam's central line \cite{berdichevsky2009variational}, as evidenced by the absence of $w_{;\alpha}$ in $W_\perp$ and $w_{(\alpha;\beta)}$ in $W_\angle $. The functionals \eqref{eq:9} and \eqref{eq:10} represent the transverse and shear energy densities, integrated over the cross section of the beam. They are positive definite and convex, so the existence of their minimizers $\check{w}_\alpha ,\check{w}$ is guaranteed. 

Denote the minima of functionals \eqref{eq:9} and \eqref{eq:10} by $\Phi_\perp (\gamma, \Omega_\alpha )$ and $\Phi_\angle (\Omega)$, respectively. Due to the linear dependence of the minimizer of \eqref{eq:9} on $\gamma $ and $\Omega _\alpha$, the function $\Phi_\perp (\gamma, \Omega_\alpha )$ is necessarily a quadratic form in these variables:
\begin{equation}
\label{eq:11}
\Phi_\perp (\gamma, \Omega_\alpha )=\frac{1}{2}( E_\perp \gamma^2+ 2E_{\perp \alpha} \gamma \Omega_{\alpha }+E_{\perp \alpha \beta } \Omega_\alpha \Omega_\beta ).
\end{equation}
Similarly, the minimizer of \eqref{eq:10} depends linearly on $\Omega$, implying that $\Phi_\angle (\Omega)$ is a quadratic function of $\Omega$:
\begin{equation}
\label{eq:12}
\Phi_\angle (\Omega)=\frac{1}{2}C_\angle \Omega^2,
\end{equation}
where $C_\angle$ represents the torsional stiffness of the beam. 

These variational problems, along with their dual formulations, can be solved numerically to determine the coefficients in \eqref{eq:11} and \eqref{eq:12} explicitly. This procedure will be detailed in the following Section. 

\section{Numerical solution of the cross-sectional problems}
\subsection{Anti-plane cross-sectional problem}
We begin with the anti-plane cross-sectional problem \eqref{eq:10}, which involves finding the warping function $w(\vb{y})$ under the constraint $\langle w\rangle =0$. This constraint, in the context of the Ansatz \eqref{eq:5} in Section 3, ensures that $v(x)$ represents the mean longitudinal displacement. Note that the constraint does not affect the minimum value of the functional \eqref{eq:10}, as the latter is invariant with respect to a constant shift in $w(\vb{y})$.

Varying the functional \eqref{eq:10} and equating its first variation to zero yields the following elliptic equation with variable coefficients:
\begin{equation}
\label{eq:13}
(\mu(\vb{y}) w_{;\alpha })_{;\alpha }=\mu(\vb{y})_{;\alpha } h\Omega e_{\alpha 
\beta}y_\beta ,
\end{equation}
subject to the Neumann boundary condition:
\begin{equation}
\label{eq:14}
w_{;\alpha } n_{\alpha }= h\Omega e_{\alpha \beta}y_\beta n_{\alpha },
\end{equation}
where $n_\alpha$ are the components of the outward normal vector to the boundary $\partial \bar{\mathcal{A}}$.

The calculus of variations provides a framework for deriving a dual variational problem \cite{berdichevsky2009variational,le1999vibrations}. For brevity, we omit the derivation and present the dual formulation: Maximize the functional
\begin{equation}
\label{eq:15}
h^2 \langle -p_\alpha h\Omega e_{\alpha \beta}y_\beta -\frac{1}{2\mu(\vb{y})} p_{\alpha }p_{\alpha } \rangle 
\end{equation}
among all functions $p_\alpha$ satisfying the constraints
\begin{equation}
\label{eq:16}
p_{\alpha;\alpha }=0 \quad \text{in $\bar{\mathcal{A}}$},\quad p_\alpha n_\alpha =0 \quad \text{at $\partial \bar{\mathcal{A}}$}.
\end{equation}
Physically, $w_{;\alpha}$ represents the shear strains, and their dual quantities $p_\alpha$ represent the shear stresses.  The strong duality property ensures that the maximum of \eqref{eq:15} under the constraints \eqref{eq:16} equals the minimum of \eqref{eq:10}.

To further simplify the dual problem, we introduce a stress function $\chi$ such that:
\begin{equation*}
\label{chi}
p_\alpha =e_{\alpha \beta} \chi_{;\beta }.
\end{equation*}
With this, the first constraint in \eqref{eq:16} is satisfied automatically. Substituting this equation into the boundary condition $p_\alpha n_\alpha =0$ and integrating, we find that $\chi$ is constant along the boundary. For a simply connected cross-section, we can set $\chi=0$ without loss of generality.  This allows us to rewrite the dual problem in terms of $\chi$:
\begin{equation}
\label{eq:17}
h^2 \langle -\chi_{;\alpha }h\Omega y_\alpha -\frac{1}{2\mu(\vb{y})} \chi_{;\alpha } \chi_{;\alpha } \rangle \to \max_{\chi |_{\partial \bar{\mathcal{A}}}=0}.
\end{equation}
This implies that $\chi$ satisfies the elliptic equation:
\begin{equation*}
\Bigl( \frac{1}{\mu(\vb{y})} \chi_{;\alpha }\Bigr)_{;\alpha }=2h\Omega ,
\end{equation*}
with the Dirichlet boundary condition $\chi=0$ at $\partial \bar{\mathcal{A}}$.

The minimum of \eqref{eq:10} and the maximum of \eqref{eq:17} can be expressed as:
\begin{equation*}
\Phi_\angle (\Omega)=\frac{1}{2} C_\angle \Omega ^2=\frac{1}{2} h^4 c \Omega ^2, 
\end{equation*}
where $C_\angle = h^4 c$ represents the torsional stiffness of the beam, and $c$ is a constant that can be computed from either the primal or dual formulation:
\begin{equation}
\label{eq:18}
\begin{split}
c &=\min_{\langle w\rangle =0} \langle \mu(\vb{y}) (w_{;\alpha }-e_{\alpha 
\beta}y_\beta)(w_{;\alpha }-e_{\alpha \gamma}y_\gamma)\rangle 
\\
& =\max_{\chi |_{\partial \bar{\mathcal{A}}}=0} \langle -2\chi_{;\alpha }y_\alpha -\frac{1}{\mu(\vb{y})} \chi_{;\alpha } \chi_{;\alpha } \rangle 
\end{split}
\end{equation}
These dual formulations provide upper and lower bounds for $c$, enabling us to estimate the accuracy of our numerical solutions.

While the dual variational problems \eqref{eq:18} generally require numerical solution, an analytical solution exists for a circular cross-section with a radially varying shear modulus, $\mu(r)$. In this case, equation \eqref{eq:13} with $h\Omega = 1$ simplifies to:
\begin{equation*}
(\mu(r) w_{;\alpha })_{;\alpha }=\mu(r)_{;r} y_\alpha e_{\alpha 
\beta}y_\beta =0,
\end{equation*}
and the boundary condition \eqref{eq:14} becomes:
\begin{equation*}
\pdv{w}{r}\Bigr|_{r=1}=e_{\alpha \beta}y_\beta y_\alpha =0.
\end{equation*}
This leads to the trivial solution $w = 0$, and the torsional stiffness constant can be expressed in closed form as:
\begin{equation*}
c=\langle \mu(r) r^2 \rangle=2\pi \int_0^1 \mu(r) r^3 \dd{r}.
\end{equation*}

To demonstrate the effectiveness of our approach, we now solve the variational problems \eqref{eq:18} numerically for a beam with a rectangular cross-section. The normalized cross-section occupies the rectangle $(-a/2, a/2) \times (-1/2, 1/2)$ in the $(y_1, y_2)$-plane, with width $a$ and height 1, where we assume $a < 1$. 
We consider a functionally graded material with effective Young's modulus $E$ and Poisson's ratio $\nu$ varying along $y_2$ according to the following power-law relationship \cite{reddy1999axisymmetric}:
\begin{equation}
\label{eq:19}
\begin{split}
&E(y_2)=E_L (1/2-y_2)^\delta +E_R (1-(1/2-y_2)^\delta ),
\\
&\nu (y_2)=\nu_L (1/2-y_2)^\delta +\nu_R (1-(1/2-y_2)^\delta ),
\end{split}
\end{equation}
where $\delta$ is the gradient index controlling the material gradation, $E_L$ and $\nu_L$ are the values on the left side ($y_2 = -1/2$), and $E_R$ and $\nu_R$ are the values on the right side ($y_2 = 1/2$), indicating that the material at $y_2 = \pm 1/2$ corresponds to one of the pure phases. This power-law model has been shown to effectively capture the material properties of two-phase functionally graded materials, as demonstrated by the theoretical and experimental results in \cite{yin2004micromechanics}. The Lamé's moduli, $\lambda$ and $\mu$, are expressed in terms of Young's modulus $E$ and Poisson's ratio $\nu$ as follows:
\begin{equation}
\label{eq:20}
\lambda = \frac{E\nu }{(1+\nu)(1-2\nu)},\quad \mu =\frac{E}{2(1+\nu)}.
\end{equation}
To facilitate numerical simulations, we normalize Young's modulus by its value on the right side, $E_R$:
\begin{equation*}
E (y_2)=E_R \bar{E}(y_2),\quad \bar{E}(y_2)=\kappa (1/2-y_2)^\delta + (1-(1/2-y_2)^\delta ),
\end{equation*}
where $\kappa = E_L / E_R$ is the ratio of Young's moduli of phases.  Consequently, the Lamé moduli are also normalized:
\begin{equation}
\begin{split}
\label{eq:21}
\lambda &= E_R \bar{\lambda}(y_2),\quad \bar{\lambda}=\frac{\bar{E}\nu }{(1+\nu)(1-2\nu)},
\\
\mu &=E_R \bar{\mu}(y_2),\quad \bar{\mu}=\frac{\bar{E}}{2(1+\nu)}.
\end{split}
\end{equation} 

The torsional stiffness constant can then be expressed as $c = E_R \bar{c}$, where $\bar{c}$ is determined from the normalized dual variational problems:
\begin{equation}
\label{eq:22}
\begin{split}
\bar{c} &=\min_{\langle w\rangle =0} \langle \bar{\mu}(y_2) (w_{;\alpha }-e_{\alpha 
\beta}y_\beta)(w_{;\alpha }-e_{\alpha \gamma}y_\gamma)\rangle 
\\
& =\max_{\chi |_{\partial \bar{\mathcal{A}}}=0} \langle -2\chi_{;\alpha }y_\alpha -\frac{1}{\bar{\mu}(y_2)} \chi_{;\alpha } \chi_{;\alpha } \rangle 
\end{split}
\end{equation}
For definiteness of the numerical analysis we assume that $\kappa<1$. 

We solve the dual variational problems \eqref{eq:22} for $\bar{c}$ in two steps. First, we address the minimization problem by solving the elliptic equation:
\begin{equation*}
(\bar{\mu}(y_2) w_{;\alpha })_{;\alpha }=-\bar{\mu}(y_2)_{;2} y_1,
\end{equation*}
subject to the Neumann boundary condition:
\begin{equation}
\label{eq:23}
\bar{\mu}(y_2) w_{;\alpha } n_{\alpha }= \bar{\mu}(y_2) e_{\alpha \beta}y_\beta n_{\alpha }.
\end{equation}
We use the finite element solvers implemented in the Matlab PDE Toolbox \cite{mathworks1995partial} to solve this boundary value problem. Note that the factor $\bar{\mu}(y_2)$ is retained on both sides of equation \eqref{eq:23} to ensure consistency with the Neumann boundary condition definition in the PDE Toolbox.  Once the solution for $w(\vb{y})$ is obtained, we compute the value of the primal functional using the `integral2` function in Matlab.

Similarly, for the maximization problem, we solve the elliptic equation:
\begin{equation*}
\Bigl( \frac{1}{\bar{\mu}(y_2)} \chi_{;\alpha }\Bigr)_{;\alpha }=2,
\end{equation*}
subject to the Dirichlet boundary condition:
\begin{equation*}
\chi \Bigr|_{\partial \bar{\mathcal{A}}}= 0.
\end{equation*}
Again, we use the PDE Toolbox to obtain the solution for $\chi(\vb{y})$ and then compute the value of the dual functional using the `integral2` function.

The primal and dual functionals provide upper and lower bounds for the true value of $\bar{c}$. The computation terminates when the difference between these bounds falls below a predefined absolute or relative tolerance, ensuring convergence to the accurate solution. The Matlab codes for solving these anti-plane cross-sectional problems are provided in Appendix \ref{A1}. These codes are flexible and can be easily extended to handle arbitrary cross-sectional shapes and material distributions.

\begin{figure}[!htb]
\centering
\includegraphics[width=0.49\textwidth]{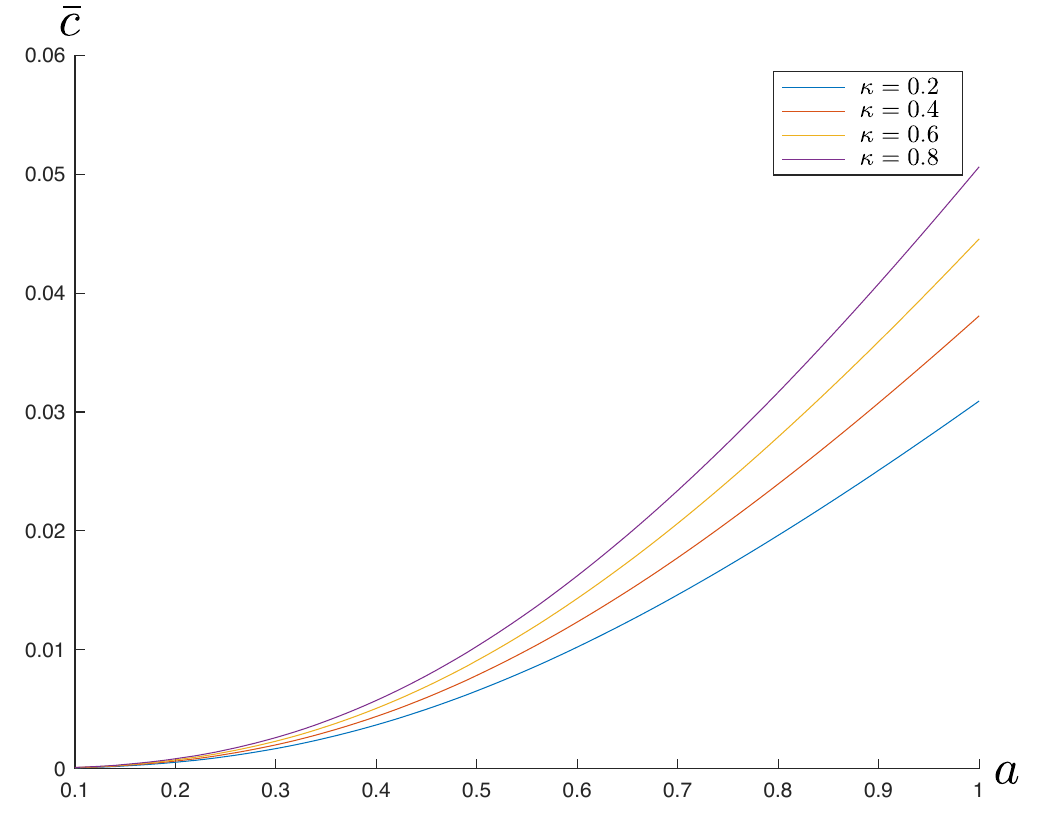}
\includegraphics[width=0.49\textwidth]{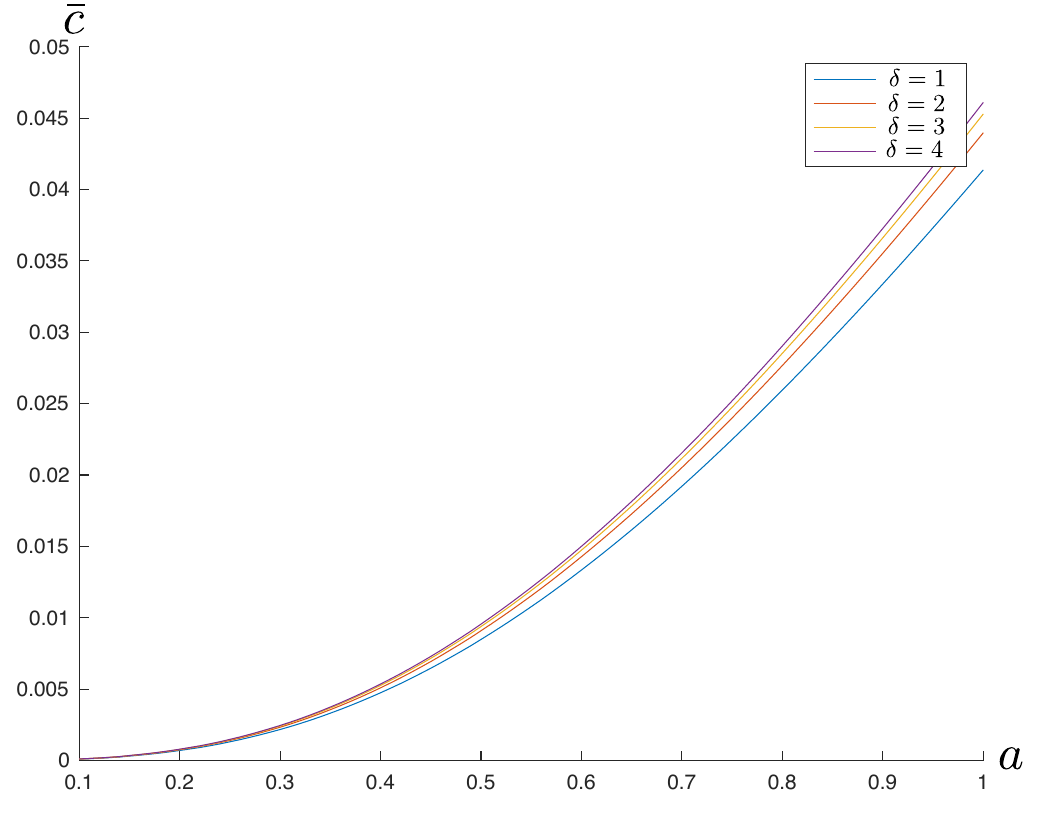} 
\caption{Dependence of normalized torsional stiffness $\bar{c}$ on the normalized width $a$: (i) at fixed $\delta=1$, $\nu_L=0.1$, $\nu_R=0.4$ and four different $\kappa=0.2,0.4,0.6,0.8$ (left), and (ii) at fixed $\kappa =0.5$, $\nu_L=0.1$, $\nu_R=0.4$ and four different $\delta=1,2,3,4$ (right).}
\label{fig:2}
\end{figure}

Figure \ref{fig:2} shows the dependence of the normalized torsional stiffness $\bar{c}$ on the normalized width $a$ of the rectangular cross-section.  In the left plot, we fix the gradient index $\delta = 1$, Poisson's ratios $\nu_L=0.1$, $\nu_R=0.4$, and vary the moduli ratio $\kappa = 0.2, 0.4, 0.6, 0.8$.  We observe that $\bar{c}(a)$ increases monotonically with $a$, indicating that wider beams exhibit higher torsional stiffness.  Furthermore, larger values of $\kappa$ (i.e., less variation in Young's modulus) lead to higher stiffness.

The right plot in Figure \ref{fig:2} presents the results for fixed $\kappa = 0.5$, $\nu_L=0.1$, $\nu_R=0.4$, and varying gradient index $\delta = 1, 2, 3, 4$.  Similar to the previous case, $\bar{c}(a)$ increases monotonically with $a$.  Additionally, we find that larger values of $\delta$ (i.e., a steeper gradient in Young's modulus) result in higher torsional stiffness.

\begin{figure}[!htb]
\centering
\includegraphics[width=0.65\textwidth]{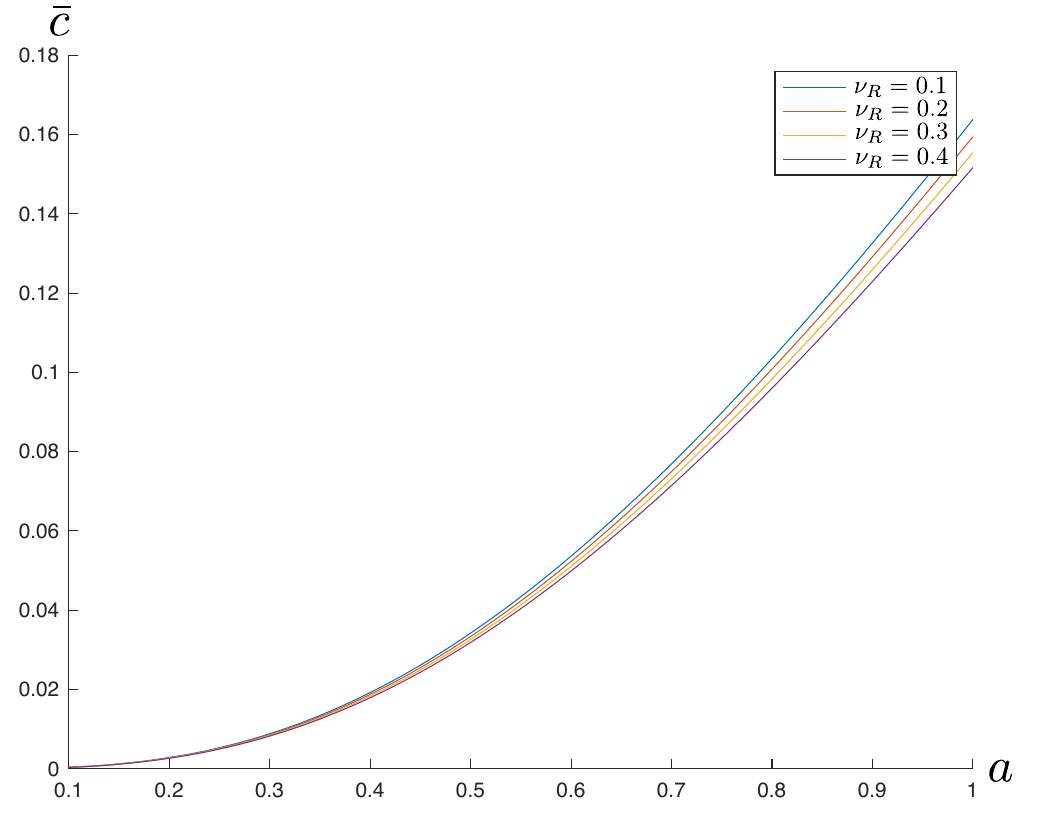}
\caption{Normalized torsional stiffness $\bar{c}$ as a function of the normalized width $a$ for fixed $\delta=0.5$, $\kappa=4$, $\nu_L=0.3$, and varying $\nu_R = 0.1, 0.2, 0.3, 0.4$.}
\label{fig:3}
\end{figure}
Figure \ref{fig:3} illustrates the influence of varying the right-side Poisson's ratio $\nu_R$ on the normalized torsional stiffness $\bar{c}$ for a fixed gradient index ($\delta = 0.5$), Young's moduli ratio ($\kappa = 4$), and left-side Poisson's ratio ($\nu_R = 0.3$).  As in the previous cases, the torsional stiffness increases monotonically with the normalized width $a$. However, we observe that increasing $\nu_R$ leads to a decrease in the torsional stiffness. This highlights the significant role of Poisson's ratio variation in the torsional response of functionally graded beams.

\subsection{Plane strain cross-sectional problem}

We now delve into the more intricate plane strain cross-sectional problem \eqref{eq:9}. The first variation of \eqref{eq:9} yields the following governing equations:
\begin{equation}
\label{eq:24}
-(\lambda(\vb{y})w_{\beta;\beta})_{;\alpha}-2(\mu(\vb{y})w_{(\alpha;\beta)})_{;\beta}=(\lambda(\vb{y})(\gamma +h\Omega_\beta y_\beta ))_{;\alpha },
\end{equation} 
and boundary conditions at $\partial \bar{\mathcal{A}}$:
\begin{equation}
\label{eq:25}
\lambda(\vb{y})w_{\beta;\beta}n_{\alpha}+2\mu(\vb{y})w_{(\alpha;\beta)}n_{\beta}=-\lambda(\vb{y})(\gamma +h\Omega_\beta y_\beta ))n_{\alpha }.
\end{equation}
These equations, along with the constraints \eqref{eq:6} that prevent rigid body motion, form a boundary-value problem analogous to that of linear plane strain elasticity for inhomogeneous materials \cite{muskhelishvili2013some}.  The right-hand sides of \eqref{eq:24} and \eqref{eq:25} act as body forces and surface tractions, respectively.

Applying the calculus of variations \cite{berdichevsky2009variational,le1999vibrations}, we derive the following dual variational problem: Maximize the functional
\begin{equation}
\label{eq:26}
I_\perp^*[\sigma_{\alpha \beta}]=h^2 \langle \nu (\vb{y}) \sigma_{\alpha \alpha}(\gamma +h\Omega _\beta y_\beta )-W_\perp^* (\sigma_{\alpha \beta}) \rangle 
\end{equation}
among all stress fields $\sigma_{\alpha \beta}$ satisfying the constraints:
\begin{equation}
\label{eq:27}
\sigma_{\alpha \beta ; \beta}=0\quad \text{in $\bar{\mathcal{A}}$}, \quad \sigma_{\alpha \beta}n_\beta =0 \quad \text{on $\partial \bar{\mathcal{A}}$}. 
\end{equation}
Here, $W_\perp^* (\sigma_{\alpha \beta})$ is the complementary energy density, given by:
\begin{equation*}
W_\perp^* (\sigma_{\alpha \beta})=\frac{1}{4\mu (\vb{y})}\sigma_{\alpha \beta}\sigma_{\alpha \beta}-\frac{\nu (\vb{y})}{4\mu (\vb{y})}(\sigma _{\lambda \lambda})^2.
\end{equation*}
The strong duality property holds in this case as well, ensuring that the maximum of \eqref{eq:26} under the constraints \eqref{eq:27} is equal to the minimum of \eqref{eq:9}.

To simplify the dual problem, we introduce the Airy stress function $\chi$:
\begin{equation*}
\sigma_{\alpha \beta}=e_{\alpha \gamma}e_{\beta \delta}\chi_{;\gamma \delta}.
\end{equation*}
With this, the equilibrium equations in \eqref{eq:27} are satisfied automatically. Substituting this into the traction free boundary conditions $\sigma_{\alpha \beta}n_\beta =0$ and integrating, we obtain the following boundary conditions for $\chi$:
\begin{equation}\label{eq:28}
\dv{\chi_{;\alpha }}{s}=0.
\end{equation}
If the stress field is known, the Airy stress function can be determined uniquely up to a linear function of $\vb{y}$. We can eliminate this arbitrariness by fixing $\chi$ and its first derivatives $\chi_{;\alpha }$ at some point $\vb{y}_0$ on the boundary $\partial \bar{\mathcal{A}}$
\begin{equation*}
\chi =\chi_{;1}=\chi_{;2}=0\quad \text{at $\vb{y}_0\in \partial \bar{\mathcal{A}}$}.
\end{equation*}
Provided the domain occupied by the cross section is simply connected, Eqs.~\eqref{eq:28} can now be integrated along the boundary to yield
\begin{equation}
\label{eq:29}
\chi |_{\partial \bar{\mathcal{A}}}=0, \quad \pdv{\chi}{n}\Bigr|_{\partial \bar{\mathcal{A}}}=0.
\end{equation}
This allows us to reformulate the dual problem in terms of $\chi$:
\begin{equation}
\label{eq:30}
h^2 \Bigl\langle \nu(\vb{y}) \laplacian \chi (\gamma + h\Omega _\alpha y_\alpha )-\frac{1}{4\mu(\vb{y})} \chi_{;\alpha \beta} \chi_{;\alpha \beta} +\frac{\nu (\vb{y})}{4\mu (\vb{y})} (\laplacian \chi )^2 \Bigr\rangle \to \max_{\chi \in \eqref{eq:29}},
\end{equation}
where $\laplacian$ denotes the Laplace operator and $\chi \in \eqref{eq:29}$ under $\max$ means that the maximum must be sought among the Airy stress functions that satisfy the boundary conditions \eqref{eq:29}. 
Varying the functional \eqref{eq:30} leads to a fourth-order partial differential equation:
\begin{equation}
\label{eq:31}
-\Bigl( \frac{1}{2\mu (\vb{y})}\chi_{;\alpha \beta}\Bigr) _{;\alpha \beta}+ \laplacian \Bigl( \frac{\nu (\vb{y})}{2\mu (\vb{y})}\laplacian \chi \Bigr)=-\laplacian (\nu(\vb{y})(\gamma + h\Omega _\alpha y_\alpha ))
\end{equation}
subject to the boundary conditions \eqref{eq:29}. This equation simplifies to the biharmonic equation when $\mu$ and $\nu$ are constant.

To facilitate the numerical solution with the Matlab PDE Toolbox, which is designed for second-order equations, we introduce the stress vector $\psi_\alpha = \chi_{;\alpha}$. The consistency condition requires that $e_{\alpha \beta }\psi_{\alpha ;\beta }=0$. In addition, the Dirichlet conditions $\psi_\alpha =0$ must be fulfilled at the edges. Note that these conditions imply the traction-free boundary conditions $\sigma_{\alpha \beta}n_\beta=0$. Indeed, using the definition of the stress vector and the identity $e_{\beta \delta}n_\beta =\tau_\delta$ results in
\begin{equation*}
\sigma_{\alpha \beta}n_\beta =e_{\alpha \gamma}e_{\beta \delta}\psi_{\gamma;\delta}n_\beta =e_{\alpha \gamma} \psi_{\gamma;\delta} \tau_\delta =e_{\alpha \gamma} \dv{\psi_\gamma }{s}=0.
\end{equation*}
Thus, $\psi_\alpha$ must be constrained by
\begin{equation}
\label{eq:32}
e_{\alpha \beta }\psi_{\alpha ;\beta }=0 \quad \text{in $\bar{\mathcal{A}}$},\quad \psi_\alpha=0 \quad \text{at $\partial \bar{\mathcal{A}}$}. 
\end{equation}
This allows us to rewrite the dual problem \eqref{eq:26} as:
\begin{equation*}
h^2 \Bigl\langle \nu(\vb{y}) \psi_{\alpha ;\alpha} (\gamma + h\Omega _\beta y_\beta )-\frac{1}{4\mu(\vb{y})} \psi_{(\alpha ;\beta)} \psi_{(\alpha ;\beta)} +\frac{\nu (\vb{y})}{4\mu (\vb{y})} (\psi_{\alpha;\alpha} )^2 \Bigr\rangle \to \max_{\psi_\alpha \in \eqref{eq:32}},
\end{equation*}
where the maximum is sought among $\psi_\alpha$ satisfying constraints \eqref{eq:32}. We enforce the constraint \eqref{eq:32}$_1$ using a penalty term, leading to the unconstrained maximization problem:
\begin{equation}
\label{eq:33}
h^2 \Bigl\langle  \nu(\vb{y}) \psi_{\alpha ;\alpha} (\gamma + h\Omega _\beta y_\beta )-\frac{1}{4\mu(\vb{y})} \psi_{(\alpha ;\beta)} \psi_{(\alpha ;\beta)} +\frac{\nu (\vb{y})}{4\mu (\vb{y})} (\psi_{\alpha;\alpha} )^2 -\theta (e_{\alpha \beta}\psi_{\alpha;\beta})^2 \Bigr\rangle ,
\end{equation}
where $\theta$ is the penalty parameter. We maximize \eqref{eq:33} among all functions $\psi_\alpha$ satisfying Dirichlet's boundary conditions \eqref{eq:32}$_2$.

Berdichevsky \cite{berdichevskii1981energy} established two important properties of this plane strain cross-sectional problem.  First, the minimum of the functional \eqref{eq:9} is zero when the Poisson's ratio of the FG beam is constant.  This can be shown by constructing functions $w_\alpha(\vb{y})$ that satisfy the constraints \eqref{eq:6} and make the functional vanish. Indeed, it is easy to check that
\begin{equation*}
w_\alpha =-\nu \gamma y_\alpha-\frac{1}{2}\nu h(\delta_{\alpha \beta}\Omega_\gamma+\delta_{\alpha \gamma}\Omega_\beta -\delta_{\beta \gamma}\Omega_\alpha) \Bigl( y_\beta y_\gamma -\frac{\langle y_\beta y_\gamma \rangle }{|\bar{\mathcal{A}}|}\Bigr) ,
\end{equation*}
with $|\bar{\mathcal{A}}|$ being the area of $\bar{\mathcal{A}}$, satisfy constraints \eqref{eq:6} and
\begin{equation*}
w_{(\alpha ;\beta )} +\nu \delta _{\alpha \beta }(\gamma +h\Omega _\gamma y_\gamma )=0.
\end{equation*}
Consequently functional \eqref{eq:9} vanishes on them. Since the functional \eqref{eq:9} is positive definite, its minimum must be zero.  This property can also be verified using the dual variational problem \eqref{eq:30}. When $\nu$ is constant, the right-hand side of equation \eqref{eq:31} vanishes, leading to the trivial solution $\chi = 0$ for the Airy stress function.  Consequently, the maximum of the dual functional \eqref{eq:30} is zero, which, by strong duality, implies a vanishing minimum for the primal functional \eqref{eq:9}. Second, for beams with central-symmetric cross-sections and elastic moduli, the coupling term between extension and bending in the average transverse energy \eqref{eq:11} vanishes, i.e., $E_{\perp \alpha} = 0$.  We will use these properties in the next Section to derive the one-dimensional beam theory.

We now demonstrate how to utilize the numerical solution of the plane strain problem to compute the stiffnesses associated with the average transverse energy.  As before, we consider a rectangular cross-section occupying the domain $(-a/2, a/2) \times (-1/2, 1/2)$ in the $(y_1, y_2)$-plane and a functionally graded material with Young's modulus $E$ and Poisson's ratio $\nu$ varying according to the power-law relationship \eqref{eq:19}. The Lamé moduli, $\lambda$ and $\mu$, are related to $E$ and $\nu$ through equations \eqref{eq:20}.  For numerical convenience, we normalize Young's modulus by its value on the right side, $E_R$, leading to the normalized Lamé moduli given by equations \eqref{eq:21}.

To illustrate the procedure, let us focus on the case of pure bending about the $x_1$-axis, where $\gamma = \Omega_1 = 0$ and $\Omega_2 \neq 0$. The bending stiffness associated with the average transverse energy can be expressed as $E_{\perp 22} = E_R h^4 \bar{e}_{\perp 22}$, where the normalized stiffness $\bar{e}_{\perp 22}$ is determined from the following normalized variational problems:
\begin{equation}
\label{eq:34}
\begin{split}
\bar{e}_{\perp 22}&=\min_{w_\alpha \in \eqref{eq:6}} \langle \bar{\lambda} (w_{\alpha ;\alpha } 
+2\nu y_2 )^2 
+2\bar{\mu }(w_{(\alpha ;\beta )} +\nu \delta _{\alpha \beta } y_2 )(w_{(\alpha ;\beta )} +\nu \delta _{\alpha \beta }y_2 )\rangle 
\\
&=\max_{\psi_\alpha|_{\partial \bar{\mathcal{A}}}=0} \langle  2\nu \psi_{\alpha ;\alpha} y_2-\frac{1}{2\bar{\mu}} \psi_{(\alpha ;\beta)} \psi_{(\alpha ;\beta)} +\frac{\nu }{2\bar{\mu }} (\psi_{\alpha;\alpha} )^2 -2\theta (e_{\alpha \beta}\psi_{\alpha;\beta})^2 \rangle
\end{split}
\end{equation} 

The primal problem in \eqref{eq:34} leads to the following system of equations:
\begin{equation*}
\begin{split}
&-(\bar{\lambda}w_{\beta;\beta})_{;\alpha}-2(\bar{\mu}w_{(\alpha;\beta)})_{;\beta}=(\bar{\lambda}y_2 )_{;\alpha },
\\
&\bar{\lambda}w_{\beta;\beta}n_{\alpha}+2\bar{\mu}w_{(\alpha;\beta)}n_{\beta}=-\bar{\lambda} y_2n_{\alpha },
\end{split}
\end{equation*}
which is solved using the finite element solver in the Matlab PDE Toolbox.  The corresponding $c$-matrix and $f$-vector characterizing this system are:
\begin{equation*}
c=\begin{pmatrix}
 2\bar{\mu} +\bar{\lambda}  & 0 & 0 & \bar{\lambda} \\
 0 & \bar{\mu} & \bar{\mu} & 0 \\
 0 & \bar{\mu} & \bar{\mu} & 0 \\
 \bar{\lambda} & 0 & 0 & 2\bar{\mu} +\bar{\lambda}
\end{pmatrix}, \quad f=\begin{pmatrix}
   (\bar{\lambda}(\gamma+h\Omega_\beta y_\beta))_{;1}   \\
   (\bar{\lambda}(\gamma+h\Omega_\beta y_\beta))_{;2}  
\end{pmatrix}
\end{equation*}
Once the solution for $w_\alpha$ is obtained, we evaluate the primal functional in \eqref{eq:34} using the `integral2` function in Matlab.

For the dual problem in \eqref{eq:34}, we solve the following system of equations:
\begin{equation}\label{eq:35}
\Bigl( \frac{1}{2\bar{\mu }}\psi_{(\alpha ;\beta)}\Bigr) _{;\beta}- \Bigl( \frac{\nu }{2\bar{\mu }}\psi _{\beta;\beta} \Bigr) _{;\alpha }- (\nu (\gamma + h\Omega _\beta y_\beta ))_{;\alpha }+2\theta e_{\alpha \beta}e_{\gamma \delta}\psi_{\gamma ;\delta \beta }=0,
\end{equation}
with Dirichlet boundary conditions. The corresponding $c$-matrix and $f$-vector are:
\begin{equation*}
c=\begin{pmatrix}
\frac{1-\nu }{2\bar{\mu}} & 0 & 0 & -\frac{\nu }{2\bar{\mu}} \\
 0 & \frac{1}{4\bar{\mu}}+2\theta & \frac{1}{4\bar{\mu}}-2\theta  & 0 \\
 0 & \frac{1}{4\bar{\mu}}-2\theta & \frac{1}{4\bar{\mu}}+2\theta & 0 \\
 -\frac{\nu }{2\bar{\mu}} & 0 & 0 & \frac{1-\nu }{2\bar{\mu}} 
\end{pmatrix}, \quad f=\begin{pmatrix}
   -(\nu (\gamma+h\Omega_\beta y_\beta))_{;1}   \\
   -(\nu (\gamma+h\Omega_\beta y_\beta))_{;2}  
\end{pmatrix} .
\end{equation*}
We employ the penalty method with increasing values of $\theta$ to enforce the constraint and ensure convergence to an accurate solution.  After obtaining the solution for $\psi_\alpha$, we evaluate the dual functional in \eqref{eq:34} using the `integral2` function.

By comparing the results of the primal and dual problems, we obtain upper and lower bounds for the normalized stiffness $\bar{e}_{\perp 22}$.  The computation terminates when the difference between these bounds falls below a predefined tolerance. The Matlab codes for solving these plane strain cross-sectional problems are provided in Appendix \ref{A2}.  These codes are flexible and can be easily adapted for arbitrary cross-sectional shapes and material distributions.

An efficient method for computing the stiffnesses associated with the average transverse energy involves utilizing the solution of the dual cross-sectional problem. We express the stiffnesses in terms of normalized quantities $\bar{e}_\perp$, $\bar{e}_{\perp \alpha}$, and $\bar{e}_{\perp \alpha \beta}$:
\begin{equation*}
E_\perp=E_R h^2 \bar{e}_\perp,\quad E_{\perp \alpha}=E_R h^3 \bar{e}_{\perp \alpha},\quad E_{\perp \alpha \beta}=E_R h^4 \bar{e}_{\perp \alpha \beta},
\end{equation*}
where $\bar{e}_\perp$, $\bar{e}_{\perp 11}$, and $\bar{e}_{\perp 22}$ are the normalized stiffnesses.  Let $\check{\psi}_\alpha$ be the solution of the dual variational problem \eqref{eq:33}, and let $I_\perp^*[\psi_\alpha]$ denote the functional \eqref{eq:33}.  We can then establish the following identity:
\begin{equation}
\label{eq:36}
\max_{\psi_\alpha|_{\partial \bar{\mathcal{A}}}=0}I_\perp^*[\psi_\alpha] = \frac{1}{2}h^2\langle \nu \check{\psi}_{\alpha ;\alpha } (\gamma+h\Omega_\alpha )\rangle .
\end{equation}
This identity follows from the fact that if we replace $\var{\psi}_\alpha$ by $\check{\psi}_\alpha $ in the first variation of the functional \eqref{eq:33}, then
\begin{equation*}
h^2 \Bigl\langle  \nu \check{\psi}_{\alpha ;\alpha} (\gamma + h\Omega _\beta y_\beta )-\frac{1}{2\mu} \check{\psi}_{(\alpha ;\beta)} \check{\psi}_{(\alpha ;\beta)} +\frac{\nu }{2\mu } (\check{\psi}_{\alpha;\alpha} )^2 -2\theta (e_{\alpha \beta}\check{\psi}_{\alpha;\beta})^2 \Bigr\rangle =0.
\end{equation*}

Based on the strong duality and the identity \eqref{eq:36}, we can compute the normalized stiffnesses as follows.  For the stiffness $\bar{e}_\perp$, we set $\gamma = 1$ and $\Omega_\alpha = 0$ in \eqref{eq:36}, leading to:
\begin{equation*}
\langle \nu \check{\psi}_{\alpha ;\alpha } \rangle =\bar{e}_\perp,
\end{equation*}
where $\check{\psi}_\alpha$ is obtained from the solution of \eqref{eq:35} with $\gamma + h\Omega_\alpha y_\alpha$ replaced by 1.  Similarly, for $\bar{e}_{\perp 11}$, we set $\gamma = 0$, $\Omega_1 = 1$, and $\Omega_2 = 0$, resulting in:
\begin{equation*}
\langle \bar{\nu }\check{\psi}_{\alpha ;\alpha } y_1 \rangle =\bar{e}_{\perp 11},
\end{equation*}
where $\check{\psi}_\alpha$ is the solution of \eqref{eq:35} with $\gamma + h\Omega_\alpha y_\alpha$ replaced by $y_1$.  An analogous formula holds for $\bar{e}_{\perp 22}$. For the cross stiffness $\bar{e}_{\perp 1}$, we compute:
\begin{equation*}
\langle \nu \check{\psi}_{\alpha ;\alpha } y_1 \rangle,
\end{equation*}
where $\check{\psi}_\alpha$ is the solution of \eqref{eq:35} with $\gamma + h\Omega_\alpha y_\alpha$ replaced by 1. An analogous formula holds for $\bar{e}_{\perp 2}$. Finally, the cross stiffness $\bar{e}_{\perp 12}$ is given by:
\begin{equation*}
\langle \nu \check{\psi}_{\alpha ;\alpha } y_1 \rangle,
\end{equation*}
where $\check{\psi}_\alpha$ is the solution of \eqref{eq:35} with $\gamma + h\Omega_\alpha y_\alpha$ replaced by $y_2$.

We defer a detailed discussion of these stiffnesses to the next Section, where we will combine them with the stiffnesses arising from the average longitudinal energy to obtain the complete one-dimensional beam theory.  In all cases, we utilize the corresponding dual problems to control and ensure the accuracy of this alternative approach.

\section{One-dimensional theory of FG beams}
Having solved the cross-sectional problems \eqref{eq:9} and \eqref{eq:10}, we can now construct the one-dimensional (1-D) theory for functionally graded (FG) beams.  We begin by substituting the Ansatz \eqref{eq:5}, where $w_\alpha (\vb{y},x)$ and $w(\vb{y},x)$  are the minimizers of \eqref{eq:9} and \eqref{eq:10}, respectively, into the energy functional \eqref{eq:2}.  Integrating over the cross-section and retaining only the asymptotically principal terms based on the estimation \eqref{eq:7}, we obtain the following 1-D energy functional:
\begin{equation}
\label{eq:37}
J[v(x),v_\alpha(x),\varphi(x)]=\int_0^L \Phi (\gamma,\Omega_\alpha,\Omega)\dd{x}-(fv+f_\alpha v_\alpha +q\varphi -q_\alpha v_{\alpha ,x})|_{x=L},
\end{equation}
where
\begin{equation*}
\Phi (\gamma,\Omega_\alpha,\Omega)=\Phi _\parallel (\gamma, \Omega_\alpha ) +\Phi _\perp (\gamma, \Omega_\alpha )+\Phi _\angle (\Omega),
\end{equation*}
with $\gamma$, $\Omega_\alpha$, and $\Omega$ given by \eqref{eq:8}, and the generalized forces defined as:
\begin{equation*}
f=\langle h^2t\rangle, \quad f_\alpha=\langle h^2 t_\alpha \rangle, \quad q=\langle h^3 e_{\alpha \beta}y_\alpha t_\beta \rangle,\quad q_\alpha=\langle h^3 y_\alpha t \rangle .
\end{equation*}
The 1-D energy density $\Phi (\gamma,\Omega_\alpha,\Omega)$ comprises three components: $\Phi _\parallel (\gamma, \Omega_\alpha )$, $\Phi _\perp (\gamma, \Omega_\alpha )$, and $\Phi _\angle (\Omega)$. The first component, $\Phi _\parallel (\gamma, \Omega_\alpha )$, arises from the longitudinal energy and is given by:
\begin{equation*}
\Phi _\parallel (\gamma, \Omega_\alpha )=\frac{1}{2}\langle h^2E(\vb{y})(\gamma +h\Omega_\sigma y_\sigma )^2\rangle =\frac{1}{2}(E_\parallel \gamma^2+2 E_{\parallel \alpha } \Omega_\alpha \gamma +E_{\parallel \alpha \beta } \Omega_\alpha \Omega_\beta ),
\end{equation*}
where the stiffnesses are defined as:
\begin{equation}\label{eq:38}
E_\parallel =h^2\langle E(\vb{y})\rangle ,\quad E_{\parallel \alpha }=h^3\langle E(\vb{y})y_\alpha \rangle ,\quad E_{\parallel \alpha \beta } = h^4 \langle E(\vb{y})y_\alpha y_\beta \rangle .
\end{equation}
For FG materials with elastic moduli varying according to \eqref{eq:19}, these stiffnesses can also be expressed in terms of the normalized quantities as follows
\begin{equation*}
E_\parallel =E_R h^2\bar{e}_\parallel ,\quad E_{\parallel \alpha }=E_R h^3\bar{e}_{\parallel \alpha} ,\quad E_{\parallel \alpha \beta } = E_R h^4 \bar{e}_{\parallel \alpha \beta} ,
\end{equation*}
where
\begin{equation}\label{eq:39}
\bar{e}_\parallel =\langle \bar{E}(\vb{y})\rangle , \quad \bar{e}_{\parallel \alpha} =\langle \bar{E}(\vb{y})y_\alpha \rangle , \quad \bar{e}_{\parallel \alpha \beta}= \langle \bar{E}(\vb{y})y_\alpha y_\beta \rangle .
\end{equation}
The other two components, $\Phi _\perp (\gamma, \Omega_\alpha )$ and $\Phi _\angle (\Omega)$, are obtained from the minimization problems \eqref{eq:9} and \eqref{eq:10}, respectively, and represent the contributions from the average transverse and shear energy densities. Using equations \eqref{eq:11} and \eqref{eq:12}, the total 1-D energy density can be expressed as:
\begin{equation*}
\begin{split}
\Phi &=\frac{1}{2}[(E_\parallel +E_\perp)\gamma^2+2 (E_{\parallel \alpha }+E_{\perp \alpha }) \Omega_\alpha \gamma +(E_{\parallel \alpha \beta }+E_{\perp \alpha \beta }) \Omega_\alpha \Omega_\beta ]+\frac{1}{2}C_\angle \Omega^2
\\
&=\frac{1}{2}E_R(h^2\bar{e} \gamma^2+2 h^3 \bar{e}_{\alpha }\Omega_\alpha \gamma +h^4 \bar{e}_{\alpha \beta }\Omega_\alpha \Omega_\beta +h^4\bar{c} \Omega^2),
\end{split}
\end{equation*}
where
\begin{equation*}
\bar{e}=\bar{e}_\parallel +\bar{e}_\perp, \quad \bar{e}_{\alpha}=\bar{e}_{\parallel \alpha} + \bar{e}_{\perp \alpha}, \quad \bar{e}_{\alpha \beta}=\bar{e}_{\parallel \alpha \beta}+\bar{e}_{\perp \alpha \beta}
\end{equation*}
are the total normalized stiffnesses. 

The 1-D energy density simplifies in two important cases.  First, if the Poisson's ratio of the beam is constant, then $\Phi _\perp (\gamma, \Omega_\alpha )\equiv 0$, and the energy density reduces to:
\begin{equation*}
\Phi =\frac{1}{2}E_R(h^2\bar{e}_\parallel \gamma^2+2 h^3 \bar{e}_{\parallel \alpha }\Omega_\alpha \gamma +h^4 \bar{e}_{\parallel \alpha \beta }\Omega_\alpha \Omega_\beta +h^4\bar{c} \Omega^2).
\end{equation*}
Second, if the cross-section and the elastic moduli are central-symmetric, the coupling term between extension and bending vanishes $\bar{e}_{\parallel \alpha }=0$, and the energy density becomes:
\begin{equation*}
\Phi =\frac{1}{2}E_R(h^2\bar{e} \gamma^2+h^4 \bar{e}_{\alpha \beta }\Omega_\alpha \Omega_\beta +h^4\bar{c} \Omega^2).
\end{equation*}

We now combine the results from Section 4 with the expressions for the longitudinal stiffnesses \eqref{eq:39} to compute the total stiffnesses for a rectangular cross-section with elastic moduli varying according to the power-law relationship \eqref{eq:19}.  It can be readily verified that $\bar{e}_1 = \bar{e}_{12} = 0$, leaving $\bar{e}$, $\bar{e}_2$, $\bar{e}_{11}$, and $\bar{e}_{22}$ as the remaining non-zero stiffnesses.

Our numerical analysis reveals that the contributions of the transverse stiffnesses $\bar{e}_\perp$, $\bar{e}_{\perp 2}$, and $\bar{e}_{\perp 11}$ to the total stiffnesses $\bar{e}$, $\bar{e}_2$, and $\bar{e}_{11}$, respectively, are negligibly small across the entire parameter range.  Since the longitudinal stiffnesses $\bar{e}_{\parallel}$, $\bar{e}_{\parallel 2}$, and $\bar{e}_{\parallel 11}$ depend solely on the normalized Young's modulus $\bar{E}(\vb{y})$, their computation does not require the solution of a boundary value problem.  Therefore, we focus our attention on the bending stiffness $\bar{e}_{22}$, for which the contribution of $\bar{e}_{\perp 22}$ cannot be neglected in general.

\begin{figure}[!htb]
\centering
\includegraphics[width=0.49\textwidth]{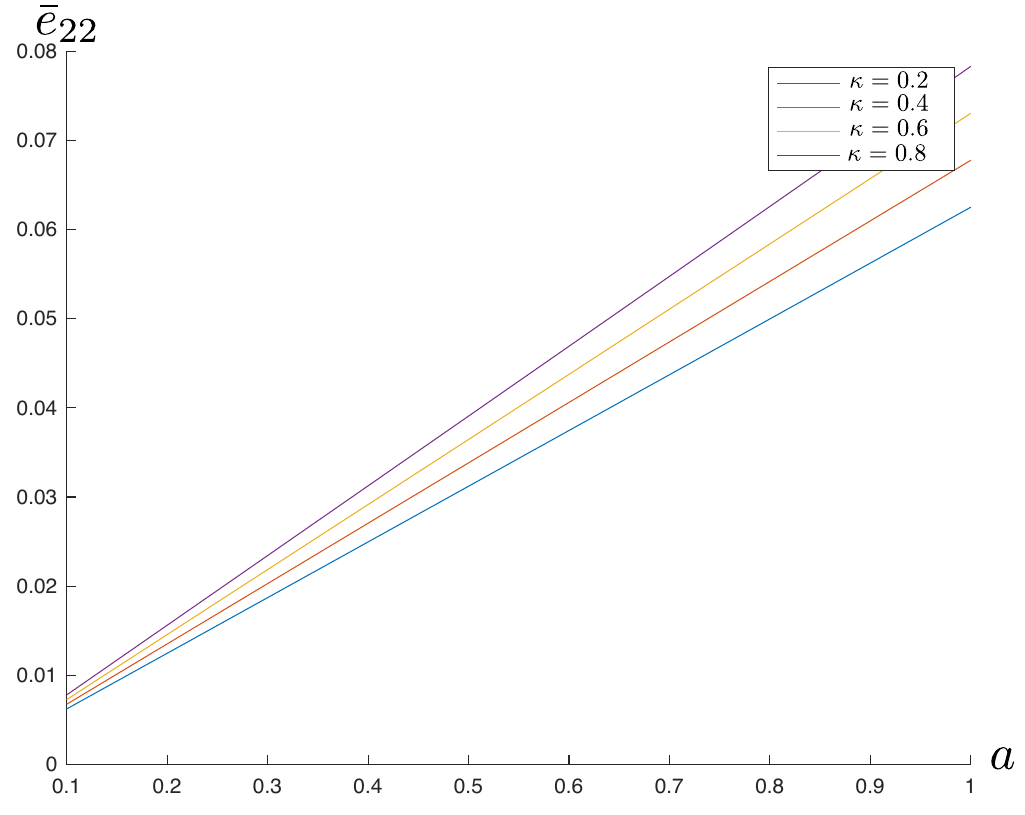}
\includegraphics[width=0.49\textwidth]{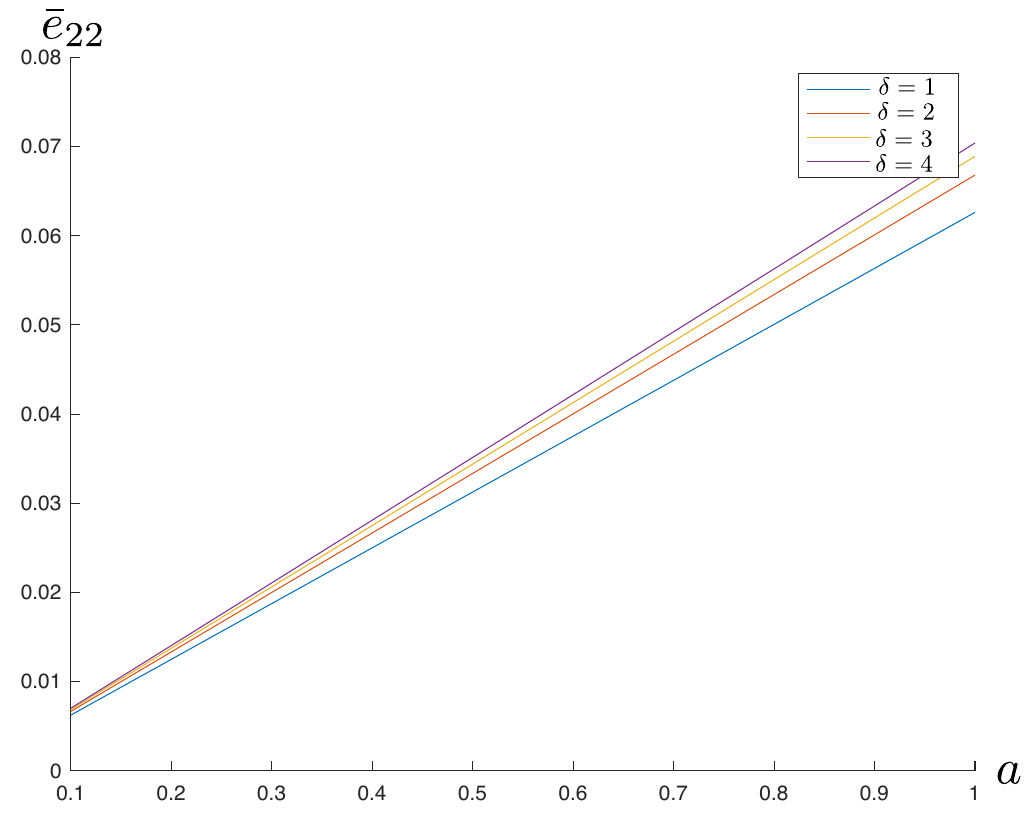} 
\caption{Normalized bending stiffness $\bar{e}_{22}$ as a function of the normalized width $a$: (a) for fixed $\delta=4$, $\nu_L=0.1$, $\nu_R=0.4$, and varying $\kappa=0.2,0.4,0.6,0.8$; (b) for fixed $\kappa =0.5$, $\nu_L=0.1$, $\nu_R=0.4$, and varying $\delta=1,2,3,4$.}
\label{fig:4}
\end{figure}

Figure \ref{fig:4} shows the variation of the normalized bending stiffness $\bar{e}_{22}$ with the normalized width $a$ of the rectangular cross-section.  In Figure \ref{fig:4} (left), we fix the gradient index $\delta = 4$ and Poisson's ratios $\nu_L=0.1$ and $\nu_R=0.4$, and vary the Young's moduli ratio $\kappa = 0.2, 0.4, 0.6, 0.8$.  We observe that $\bar{e}_{22}$ increases monotonically with $a$, indicating that wider beams exhibit higher bending stiffness. Furthermore, larger values of $\kappa$ (i.e., less variation in Young's modulus) lead to higher stiffness.

Figure \ref{fig:4} (right) presents the results for fixed $\kappa = 0.5$, $\nu_L=0.1$, $\nu_R=0.4$, and varying gradient index $\delta = 1, 2, 3, 4$.  Similar to the previous case, $\bar{e}_{22}$ increases monotonically with $a$. Additionally, we find that larger values of $\delta$ (i.e., a steeper gradient in Young's modulus) result in higher bending stiffness.

\begin{figure}[!htb]
\centering
\includegraphics[width=0.65\textwidth]{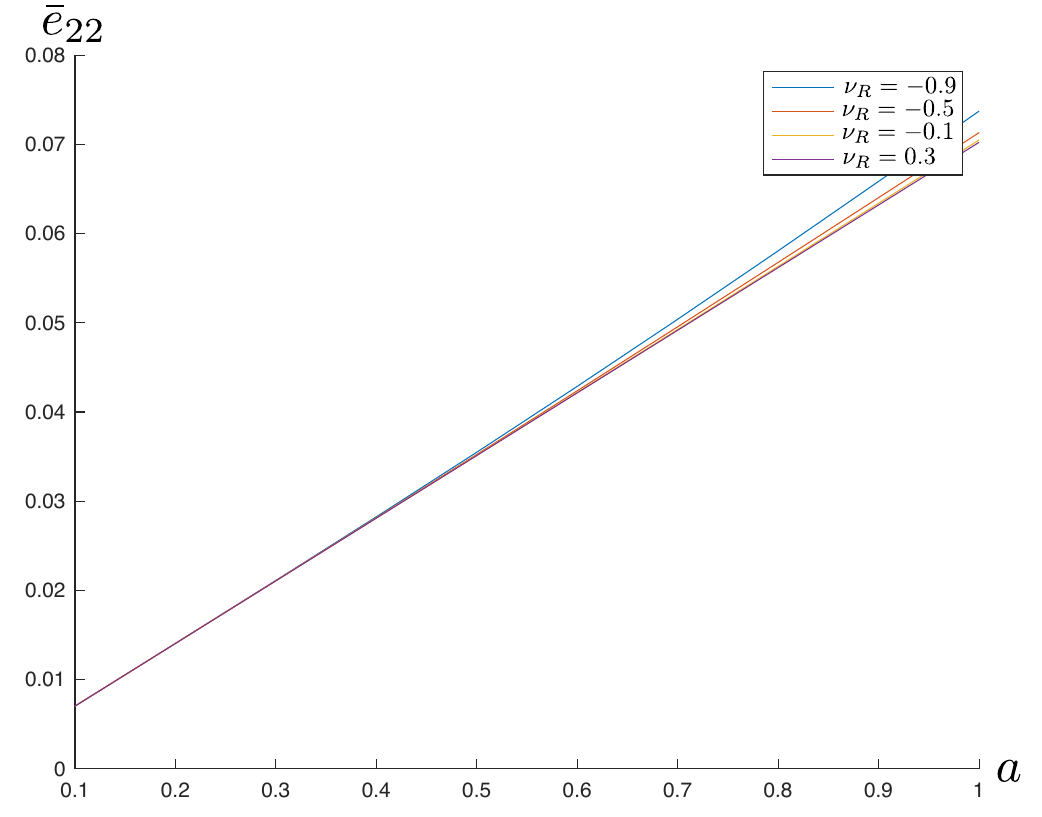}
\caption{Normalized bending stiffness $\bar{e}_{22}$ as a function of the normalized width $a$ for fixed $\delta=4$, $\kappa=0.5$, $\nu_L=0.3$, and varying $\nu_R = -0.9, -0.5, -0.1, 0.3$.}
\label{fig:5}
\end{figure}

Figure \ref{fig:5} illustrates the influence of varying the right-side Poisson's ratio $\nu_R$ on the normalized bending stiffness $\bar{e}_{22}$ for a fixed gradient index ($\delta = 4$), Young's moduli ratio ($\kappa = 0.5$), and left-side Poisson's ratio ($\nu_L = 0.3$).  We vary $\nu_R$ across its theoretically permissible range of $(-1, 0.5)$.  As in the previous cases, the bending stiffness increases monotonically with the normalized width $a$.  Since $\delta$ and $\kappa$ are fixed, the normalized Young's modulus is also fixed, implying that the longitudinal stiffness contribution $\bar{e}_{\parallel 22}$ is the same for all curves.  Therefore, the differences between the curves are solely attributed to the transverse stiffness contribution $\bar{e}_{\perp 22}$, which can be significant.

\begin{figure}[!htb]
\centering
\includegraphics[width=0.49\textwidth]{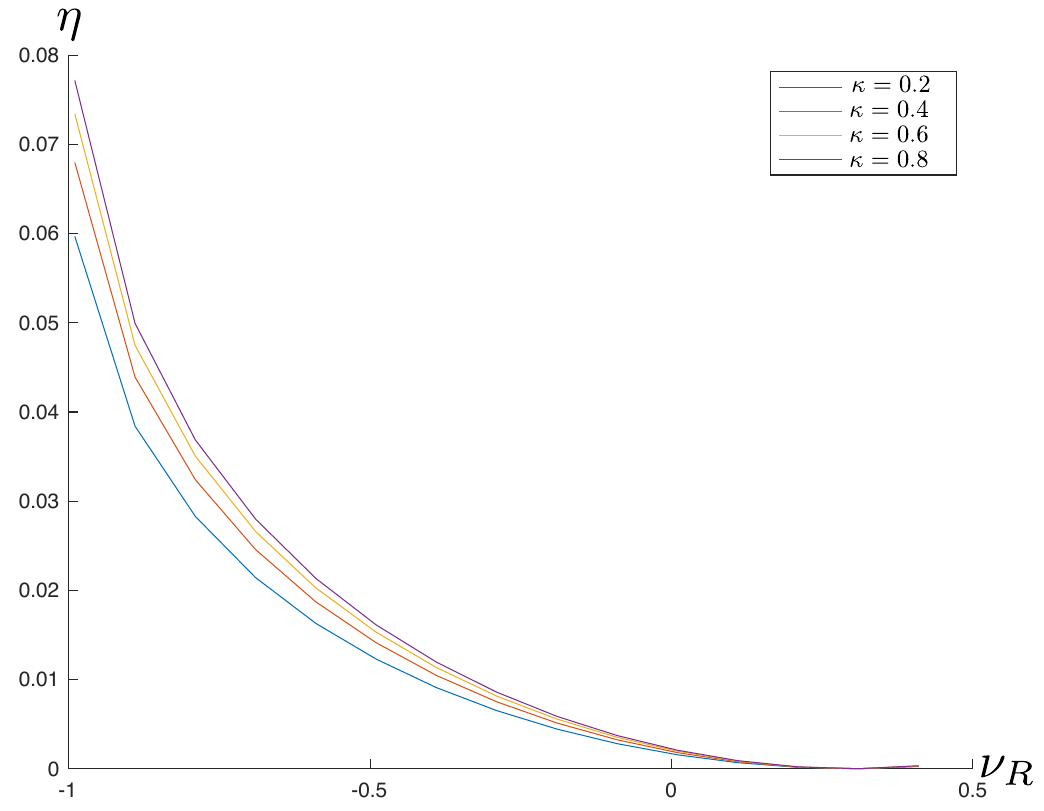}
\includegraphics[width=0.49\textwidth]{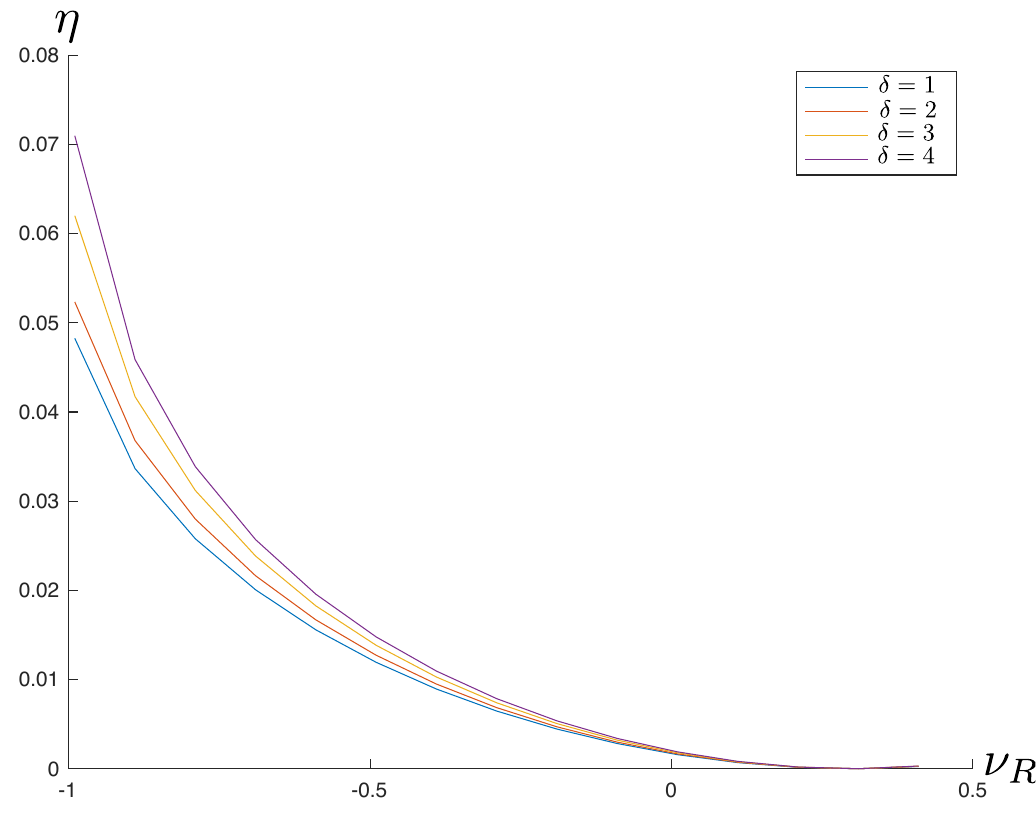} 
\caption{Percentage contribution of the transverse bending stiffness $\eta$ as a function of the right-side Poisson's ratio $\nu_R$: (a) for fixed $a=1$, $\delta=4$, $\nu_L=0.3$, and varying $\kappa=0.2,0.4,0.6,0.8$; (b) for fixed $a=1$, $\kappa =0.5$, $\nu_L=0.3$, and varying $\delta=1,2,3,4$.}
\label{fig:6}
\end{figure}

To quantify the contribution of the transverse bending stiffness to the total bending stiffness, we define the percentage $\eta = \bar{e}_{\perp 22} / (\bar{e}_{\parallel 22} + \bar{e}_{\perp 22})$.  Figure \ref{fig:6} shows the variation of $\eta$ with the right-side Poisson's ratio $\nu_R$.  In Figure \ref{fig:6} (left), we fix the normalized width $a = 1$, the gradient index $\delta = 4$, and the left-side Poisson's ratio $\nu_L = 0.3$, and vary the Young's moduli ratio $\kappa = 0.2, 0.4, 0.6, 0.8$.  As expected, $\eta$ vanishes when $\nu_R = \nu_L = 0.3$, indicating that the transverse stiffness contribution vanishes when Poisson's ratio is constant.  The percentage $\eta$ increases as $\nu_R$ approaches -1, reaching its maximum value at the lower limit of the permissible range for $\nu_R$.  This trend is similar to the observation made in \cite{le2020asymptotically} for fiber-reinforced beams.  Furthermore, larger values of $\kappa$ lead to a larger percentage contribution of the transverse stiffness.

Figure \ref{fig:6} (right) presents the results for fixed $a = 1$, $\kappa = 0.5$, $\nu_L = 0.3$, and varying gradient index $\delta = 1, 2, 3, 4$.  Similar to the previous case, $\eta$ vanishes when $\nu_R = \nu_L = 0.3$ and increases as $\nu_R$ approaches -1.  Additionally, we observe that larger values of $\delta$ result in a larger percentage contribution of the transverse stiffness.

The 1-D theory can now be formulated as a variational principle: The true average displacements $\check{v}, \check{v}_\alpha$ and the true twist angle $\check{\varphi}$ of the beam in equilibrium minimize the functional \eqref{eq:37} among all admissible displacements and twist angles that satisfy the kinematic boundary conditions:
\begin{equation}
\label{eq:40}
v(0)=v_\alpha (0)=v_{\alpha ,x}(0)=\varphi(0)=0.
\end{equation}
Conditions \eqref{eq:40} represent a clamped edge at $x = 0$.  The variational principle can be readily extended to other boundary conditions and loading scenarios \cite{le1999vibrations}.

The minimization of \eqref{eq:37} yields the following 1-D equilibrium equations:
\begin{equation}
\label{eq:41}
T_{,x}=0,\quad M_{\alpha ,xx}=0, \quad M_{,x}=0,
\end{equation}
subject to the boundary conditions \eqref{eq:40} and:
\begin{equation*}
T(L)=f,\quad M_\alpha (L)=q_\alpha,\quad M_{\alpha,x}(L)=f_\alpha ,\quad M(L)=q.
\end{equation*}
Here, $T$, $M_\alpha $, and $M$ are the tension, bending moments, and twisting moment, respectively, defined as:
\begin{equation}
\label{eq:42}
\begin{split}
&T=\pdv{\Phi}{\gamma}=(E_\parallel +E_\perp )\gamma +(E_{\parallel \alpha}+E_{\perp \alpha})\Omega_\alpha ,
\\
&M_\alpha =\pdv{\Phi}{\Omega_\alpha}=(E_{\parallel \alpha \beta }+E_{\perp \alpha \beta})\Omega_\beta +(E_{\parallel \alpha}+E_{\perp \alpha})\gamma ,
\\
&M=\pdv{\Phi}{\Omega}=C_\angle \Omega .
\end{split}
\end{equation} 
Substituting \eqref{eq:42} combined with \eqref{eq:8} into \eqref{eq:41} leads to a system of differential equations for $v$, $v_\alpha$, and $\varphi$.  Note that the equation for the twist angle $\varphi$ is uncoupled from the equations for the displacements $v$ and $v_\alpha$. This is no longer the case when we consider a curved beam.

To complete the 1-D theory, we need to establish a procedure for recovering the 3-D stress and strain fields from the 1-D solution.  Within the first-order approximation, the 3-D strain field is recovered using equations \eqref{eq:7}, where $w_\alpha(\vb{y})=\check{w}_\alpha(\vb{y})$ and $w(\vb{y})=\check{w}(\vb{y})$ are the minimizers of \eqref{eq:9} and \eqref{eq:10}, respectively. The 3-D stress field is then obtained using Hooke's law:
\begin{equation}
\label{eq:43}
\begin{split}
\sigma_{\alpha \beta}&=\lambda (\varepsilon_{\alpha \alpha}+\varepsilon_{33})+2\mu \varepsilon_{\alpha \beta}=\lambda \check{w}_{\alpha;\alpha}+2\mu \check{w}_{(\alpha ;\beta)}+\lambda (\gamma +h\Omega_\alpha y_\alpha ),
\\
\sigma_{\alpha 3}&=2\mu \varepsilon_{\alpha 3}=\mu (\check{w}_{;\alpha }-he_{\alpha \beta}\Omega y_\beta),
\\
\sigma_{33}&=\lambda (\varepsilon_{\alpha \alpha}+\varepsilon_{33})+2\mu \varepsilon_{33}=\lambda \check{w}_{\alpha;\alpha} + (\lambda+2\mu )(\gamma +h\Omega_\alpha y_\alpha),
\end{split}
\end{equation} 
where the $\vb{y}$-argument in $\lambda(\vb{y})$ and $\mu (\vb{y})$ is suppressed for short. The following relationships connect the 3-D stress components to the 1-D beam characteristics:
\begin{equation}
\label{eq:44}
\langle \sigma_{33} \rangle =\frac{T}{h^2},\quad \langle \sigma_{33} y_\alpha \rangle =\frac{M_\alpha}{h^3},\quad \langle e_{\alpha \beta}y_\alpha \sigma_{\beta 3}\rangle =\frac{M}{h^3}.
\end{equation}

To establish the connection between the 3-D stress field and the 1-D beam characteristics, we begin by proving the last relationship in equation \eqref{eq:44}.  The minimum of the functional \eqref{eq:10} can be expressed as:
\begin{equation}
\label{eq:45}
h^2 \langle \frac{1}{2}\mu(\vb{y}) (\check{w}_{;\alpha }-h\Omega e_{\alpha 
\beta}y_\beta)(\check{w}_{;\alpha }-h\Omega e_{\alpha \gamma}y_\gamma)\rangle =\frac{1}{2}C_\angle \Omega ^2,
\end{equation}
where $\check{w}$ is the minimizer of \eqref{eq:10}.  Since $\check{w}$ satisfies the Euler-Lagrange equation corresponding to \eqref{eq:10}, the following condition holds: 
\begin{equation}
\label{eq:46}
\langle \mu(\vb{y}) (\check{w}_{;\alpha }-h\Omega e_{\alpha 
\beta}y_\beta)\check{w}_{;\alpha } \rangle =0,
\end{equation}
because the left-hand side is the first variation of \eqref{eq:10} with $\delta w=\check{w}$. Substituting the expression for $\sigma_{\alpha 3}$ from \eqref{eq:43} into equation \eqref{eq:45} and using \eqref{eq:46}, we obtain:
\begin{equation*}
h^3 \langle e_{\alpha \beta}y_\alpha \sigma_{\beta 3}\rangle =C_\angle \Omega =M,
\end{equation*}
which proves the last relationship in \eqref{eq:44}. 

To prove the first two relationships in \eqref{eq:44}, we decompose the stress component $\sigma_{33}$ into two parts:
\begin{equation*}
\begin{split}
&\sigma_{33}=\pdv{W}{\varepsilon_{33}}=\sigma_{\parallel 33}+\sigma_{\perp 33},
\\
&\sigma_{\parallel 33}=\pdv{W_\parallel}{\varepsilon_{33}}=E\varepsilon_{33}=E(\gamma+h\Omega_\alpha y_\alpha ),
\\
&\sigma_{\perp 33}=\pdv{W_\perp}{\varepsilon_{33}}=\lambda (\varepsilon_{\alpha \alpha}+2\nu \varepsilon_{33})=\lambda (\check{w}_{\alpha ;\alpha }+2\nu (\gamma+h\Omega_\alpha y_\alpha )).
\end{split}
\end{equation*}
where $\sigma_{\parallel 33}$ is the stress due to the longitudinal energy density $W_\parallel$, and $\sigma_{\perp 33}$ is the stress due to the transverse energy density $W_\perp$.  Using the definitions \eqref{eq:38}, we can directly verify that:
\begin{equation}
\label{eq:47}
\begin{split}
&h^2\langle \sigma_{\parallel 33}\rangle =h^2\langle E(\gamma+h\Omega_\alpha y_\alpha ) \rangle =E_\parallel \gamma +E_{\parallel \alpha}\Omega_\alpha,
\\
&h^3\langle \sigma_{\parallel 33} y_\alpha \rangle =h^3\langle E(\gamma+h\Omega_\beta y_\beta ) y_\alpha \rangle =E_{\parallel \alpha}\gamma +E_{\parallel \alpha \beta}\Omega_\beta .
\end{split}
\end{equation}
For the stress component $\sigma_{\perp 33}$, we express the minimum of $h^2 \langle W_\perp \rangle$ as:
\begin{equation*}
\min_{w_\alpha \in \eqref{eq:6}}h^2 \langle W_\perp \rangle =\Phi_\perp (\gamma, \Omega_\alpha )=\frac{1}{2}( E_\perp \gamma^2+ 2E_{\perp \alpha} \gamma \Omega_{\alpha }+E_{\perp \alpha \beta } \Omega_\alpha \Omega_\beta ).
\end{equation*}
Using the identity \eqref{eq:4}$_2$, we can rewrite the left-hand side as:
\begin{equation*}
\min_{w_\alpha \in \eqref{eq:6}}h^2\langle W_\perp \rangle =\frac{1}{2} h^2 \langle \sigma_{\alpha \beta} \check{w}_{\alpha ;\beta}+\sigma_{33}\varepsilon_{33}-E(\varepsilon _{33})^2 \rangle .
\end{equation*}
The first term on the right-hand side, $\langle \sigma_{\alpha \beta} \check{w}_{\alpha ;\beta} \rangle$, vanishes due to the governing equation \eqref{eq:24} and boundary condition \eqref{eq:25}.  Therefore,
\begin{equation*}
\begin{split}
\frac{1}{2}h^2\langle \sigma_{\perp 33}\varepsilon_{33}\rangle &= \frac{1}{2}h^2\langle \lambda (\check{w}_{\alpha ;\alpha }+2\nu (\gamma+h\Omega_\alpha y_\alpha ))(\gamma+h\Omega_\beta y_\beta )\rangle 
\\
&=\frac{1}{2}( E_\perp \gamma^2+ 2E_{\perp \alpha} \gamma \Omega_{\alpha }+E_{\perp \alpha \beta } \Omega_\alpha \Omega_\beta ).
\end{split}
\end{equation*}
This equation holds for arbitrary $\gamma$ and $\Omega_\alpha$. To analyze the contributions of extension and bending separately, we decompose the longitudinal strain $\varepsilon_{33}$ as $\varepsilon_{33} = \varepsilon^\prime_{33} + \varepsilon^{\prime \prime}_{33}$, where $\varepsilon^\prime_{33} = \gamma$ represents the extensional component and $\varepsilon^{\prime \prime}_{33} = h\Omega_\alpha y_\alpha$ represents the bending component.  Let $\sigma^\prime_{\perp 33}$ and $\sigma^{\prime \prime}_{\perp 33}$ be the corresponding stresses induced by these strain components, respectively.  By virtue of linearity and the reciprocity relation $\langle \sigma^\prime _{\perp 33}\varepsilon^{\prime \prime}_{33}\rangle= \langle \sigma^{\prime \prime}_{\perp 33}\varepsilon^{\prime}_{33}\rangle$, we can analyze the cases of pure extension ($\gamma \neq 0, \Omega_\alpha = 0$) and pure bending ($\gamma = 0, \Omega_\alpha \neq 0$) independently to obtain the following expressions:
\begin{equation*}
\begin{split}
&h^2\langle \sigma_{\perp 33}\rangle =h^2\langle \sigma^{\prime}_{\perp 33}+\sigma^{\prime \prime}_{\perp 33}\rangle =E_\perp \gamma +E_{\perp \alpha}\Omega_\alpha,
\\
&h^3\langle \sigma_{\perp 33} y_\alpha \rangle =h^3\langle \sigma^{\prime}_{\perp 33}y_\alpha +\sigma^{\prime \prime}_{\perp 33}y_\alpha \rangle =E_{\perp \alpha}\gamma +E_{\perp \alpha \beta}\Omega_\beta .
\end{split}
\end{equation*}
Combining these results with equations \eqref{eq:47} and \eqref{eq:42} proves the first two relationships in \eqref{eq:44}.

\section{Error estimation of the one-dimensional theory}
In this Section, we utilize the Prager-Synge identity \cite{prager1947approximations} to estimate the error incurred by the 1-D FG beam theory developed in the previous Sections.

We introduce the linear space of stress fields $\vb*{\sigma}(\vb{x})$ equipped with the energetic norm:
\begin{equation*}
\norm{\vb*{\sigma}}^2_{L_2}=C_2[\vb*{\sigma}]=\int_{\mathcal{B}} W^*(\vb{x},\vb*{\sigma})\dd[3]{x},
\end{equation*}
where $W^*(\vb{x},\vb*{\sigma})$ is the positive definite complementary energy density:
\begin{equation*}
W^*(\vb{x},\vb*{\sigma})=\frac{1}{4\mu}\sigma_{ij}\sigma_{ij}-\frac{\lambda}{4\mu(3\lambda +\mu)}(\sigma_{kk})^2.
\end{equation*}
We define two types of admissible stress fields: 
\begin{enumerate}
  \item Kinematically admissible stress fields:  Stress fields $\hat{\vb*{\sigma}}$ that are derived from a displacement field $\hat{\vb{u}}$ satisfying the kinematic constraints:
\begin{equation}
\label{eq:48}
\hat{\vb*{\varepsilon}} = \frac{1}{2}(\grad \hat{\vb{u}}+(\grad \hat{\vb{u}})^T),\quad \hat{\vb{u}}=\vb{0} \quad \text{at $x=0$},
\end{equation}
where $\hat{\vb*{\varepsilon}}$ is related to $\hat{\vb*{\sigma}}$ through Hooke's law.
  \item Statically admissible stress fields: Stress fields $\tilde{\vb*{\sigma}}$ that satisfy the equilibrium conditions:
\begin{equation*}
\div \tilde{\vb*{\sigma}}=\vb{0}, \quad \tilde{\vb*{\sigma}}\vdot \vb{n}=\vb{0} \quad \text{at $\partial \mathcal{A}\times (0,L)$},\quad \tilde{\vb*{\sigma}}\vdot \vb{n}=\vb{t} \quad \text{at $x=L$}.
\end{equation*}
\end{enumerate}
Let $\check{\vb*{\sigma}}$ be the true stress state in the FG beam.  The Prager-Synge identity\footnote{For a generalization of this identity to piezoelectricity and its application in generating error estimations for two-dimensional theories of laminated and functionally graded shells, see  \cite{le1986theory,le2016asymptotically,le2017asymptotically}.} relates the true stress to any kinematically admissible stress $\hat{\vb*{\sigma}}$ and any statically admissible stress $\tilde{\vb*{\sigma}}$:
\begin{equation}
\label{eq:49}
C_2[\check{\vb*{\sigma}}-\frac{1}{2}(\hat{\vb*{\sigma}}+\tilde{\vb*{\sigma}})]=C_2[\frac{1}{2}(\hat{\vb*{\sigma}}-\tilde{\vb*{\sigma}})].
\end{equation}
This identity implies that if we can find $\hat{\vb*{\sigma}}$ and $\tilde{\vb*{\sigma}}$ such that their difference is small in the energetic norm, then $\frac{1}{2}(\hat{\vb*{\sigma}}+\tilde{\vb*{\sigma}})$ can be considered a good approximation to the true stress, and $C_2[\frac{1}{2}(\hat{\vb*{\sigma}}-\tilde{\vb*{\sigma}})]$ provides an error estimate.

Based on this identity, we can establish the following error estimation for the 1-D FG beam theory:

{\it Theorem.} The stress state determined by the 1-D theory differs from the exact 3-D stress state by a quantity of order $h/L$ in the $L_2$ norm.

To prove this theorem, we need to construct kinematically and statically admissible stress fields that are close to the stress state predicted by the 1-D theory.

{\it Kinematically admissible stress state.} We construct a kinematically admissible displacement field as:
\begin{equation}\label{eq:50}
\begin{split}
&\hat{u}_\alpha (\vb{y},x)=v_\alpha (x)-he_{\alpha \beta}\varphi (x)y_\beta+hw_\alpha (\vb{y},x),
\\
&\hat{u}(\vb{y},x)=v(x)-h v_{\alpha ,x} (x) y_\alpha +hw(\vb{y},x),
\end{split} 
\end{equation}
where $w_\alpha (\vb{y},x)$ and $w(\vb{y},x)$ are the minimizers of \eqref{eq:9} and \eqref{eq:10}, respectively, and the quantities without hats refer to the solution of the 1-D beam theory.  For simplicity, we assume regular boundary conditions at the clamped end ($x=0$), meaning that the 3-D kinematic boundary conditions agree with the inner expansion of the displacement field. This ensures that the strain field derived from \eqref{eq:50} is kinematically admissible. The proof can be extended to irregular boundary conditions by considering the additional strain field in the boundary layer, which contributes an error of order $h/L$ \cite{gregory1984decaying}.

Using the 3-D kinematic formula \eqref{eq:48}$_1$ and performing an asymptotic analysis similar to that in Section 3, we can show that the strain field $\hat{\vb*{\varepsilon}}$ differs from the strain field $\vb*{\varepsilon}$ predicted by the 1-D theory by a quantity of order $h/L$. Consequently, the corresponding stress field $\hat{\vb*{\sigma}}$, obtained from Hooke's law, also differs from the stress field $\vb*{\sigma}$ recovered from the 1-D theory according to \eqref{eq:43} by a quantity of order $h/L$.

{\it Statically admissible field.} Let us write down the exact 3-D equilibrium equations in terms of the coordinates $y_\alpha$ and $x$: 
\begin{equation}
\begin{split}
\tilde{\sigma}_{\alpha \beta ;\beta }
+h \tilde{\sigma}_{\alpha 3,x}=0,
\\
\tilde{\sigma}_{\alpha 3;\alpha}+h\tilde{\sigma}_{33,x}=0.
\end{split}
\label{eq:51}
\end{equation}
These equations are subjected to the traction-free boundary conditions
\begin{equation}
\tilde{\sigma }_{\alpha \beta }n_\beta =0,\quad
\tilde{\sigma }_{\alpha 3}n_\alpha =0 \quad \text{at $\partial \mathcal{A}\times (0,L)$.}
\label{eq:52}
\end{equation}
We construct a statically admissible stress field $\tilde{\vb*{\sigma}}$ by first setting $\tilde{\sigma}_{33}$ equal to the stress component $\sigma_{33}$ from the 1-D theory:
\begin{equation*}
\tilde{\sigma }_{33}(y_\alpha ,x)=\lambda w_{\alpha;\alpha} + (\lambda+2\mu )(\gamma (x)+h\Omega_\alpha (x) y_\alpha),
\end{equation*}
where $w_\alpha (\vb{y},x)$ is the minimizer of \eqref{eq:9}, and the quantities without tildes refer to the solution of the 1-D beam theory. We then determine the remaining stress components, $\tilde{\sigma}_{\alpha \beta}$ and $\tilde{\sigma}_{\alpha 3}$ according to
\begin{equation}
\label{eq:53}
\begin{split}
\tilde{\sigma}_{\alpha \beta}&=\lambda \tilde{w}_{\alpha;\alpha}+2\mu \tilde{w}_{(\alpha ;\beta)}+\lambda (\gamma (x)+h\Omega_\alpha (x)y_\alpha ),
\\
\tilde{\sigma}_{\alpha 3}&=\mu (\tilde{w}_{;\alpha }-he_{\alpha \beta}\Omega (x)y_\beta),
\end{split}
\end{equation} 
where $\tilde{w}_\alpha$ and $\tilde{w}$ are still unknowns. Substituting \eqref{eq:53} into the 3-D equilibrium equations \eqref{eq:51} and traction-free boundary conditions \eqref{eq:52}, we get a coupled system of  equations that differ from the equations of the plane strain and anti-plane cross-sectional problems by terms of order $h/L$. The solvability of these equations is ensured by the relationships \eqref{eq:44}. The resulting stress field $\tilde{\vb*{\sigma}}$ differs from the stress field $\vb*{\sigma}$ recovered from the 1-D theory according to \eqref{eq:43} by a quantity of order $h/L$. Note also that with the selected stress field, the boundary condition at the edge of the beam $x=L$ can only be fulfilled ``on average''. However, according to the Saint-Venant principle \cite{berdichevskii1974proof}, the additional stress field due to the self-balanced traction decreases exponentially from the edge $x=L$, so that the energy of the boundary layer is of the order of $h/L$ compared to that of the inner region. 

Therefore, we have constructed kinematically and statically admissible stress fields that are both within $O(h/L)$ of the stress field predicted by the 1-D theory.  Applying the Prager-Synge identity \eqref{eq:49}, we conclude that the error of the 1-D theory is of order $h/L$ in the energetic norm, completing the proof of the theorem.

\section{Conclusion}
This paper employed the variational-asymptotic method to construct a one-dimensional theory for functionally graded beams with arbitrary cross-sectional shapes and arbitrary material gradation. By solving the dual cross-sectional problems using a finite element implementation, we established accurate strain and stress recovery procedures. Additionally, we obtained a dual variational formulation of the plane strain problem in terms of the stress vector. An error estimate of order $h/L$ in the energetic norm, proved using the Prager-Synge identity, confirmed the accuracy of the 1D theory for slender beams. Notably, for functionally graded materials with elastic moduli varying according to \eqref{eq:19}, the average transverse energy and the corresponding stiffnesses were found to be significant in a certain range of parameters, highlighting the importance of a comprehensive cross-sectional analysis. This framework can be extended to curved functionally graded beams, large deformations, and smart materials, topics that will be addressed in our forthcoming papers.

\begin{appendices}
\section{Appendix: Matlab-codes}
This Appendix provides four Matlab codes: two for solving the primal and dual formulations of the anti-plane cross-sectional problem, and two for solving the primal and dual formulations of the plane-strain cross-sectional problem.

\subsection{Anti-plane cross-sectional problem}\label{A1}

\subsubsection{Primal formulation}
\begin{lstlisting}
clear;
clc;
% Create a model
model = createpde();
% Include a rectangle of width ar in the model and plot the geometry
ar = 1;
R1 = [3 
      4 
      -ar/2 
      ar/2 
      ar/2 
      -ar/2 
      0.5 
      0.5 
      -0.5 
      -0.5];
g = decsg(R1);
geometryFromEdges(model,g);
figure(1)
pdegplot(model,"EdgeLabels","on") 
xlim([-ar/2-0.1 ar/2+0.1]) 
ylim([-0.6 0.6])
% Generate a mesh with a maximum edge length of 0.01. Plot the mesh.
generateMesh(model,"Hmax",0.01,"GeometricOrder","quadratic"); 
figure(2) 
pdemesh(model)
% Specify the PDE coefficients
kappa=0.5;
delta=0.5;
nu_l=0.1;
nu_r=0.4;
ccoeff = @(location,state)ccoeffunction(location,state,kappa,delta,nu_l,nu_r);
fcoeff = @(location,state)fcoeffunction(location,state,kappa,delta,nu_l,nu_r);
specifyCoefficients(model,"m",0,"d",0,"c",ccoeff,"a",0,"f",fcoeff);
% Specify the boundary condition 
applyBoundaryCondition(model,"neumann", "Edge",1, "g",...
    @(location,state)bcfuncxm(location,state,kappa,delta,nu_l,nu_r)); 
applyBoundaryCondition(model,"neumann", "Edge",3, "g",...
    @(location,state)bcfuncxp(location,state,kappa,delta,nu_l,nu_r));
applyBoundaryCondition(model,"neumann", "Edge",2, "g",...
    @(location,state)bcfuncyp(location,state,kappa,delta,nu_l,nu_r)); 
applyBoundaryCondition(model,"neumann", "Edge",4, "g",...
    @(location,state)bcfuncym(location,state,kappa,delta,nu_l,nu_r));
% Solve the equation and plot the solution.
results = solvepde(model);
u = results.NodalSolution;
figure(3)
pdeplot(model,"XYData",u)
title("Numerical Solution");
xlabel("x")
ylabel("y")
% Retrieve the mesh data
p = model.Mesh.Nodes;
t = model.Mesh.Elements;
% Compute the torsional stiffness
func = @(x,y) f_integrand(x,y,results,kappa,delta,nu_l,nu_r);
torsion = integral2(func,-ar/2,ar/2,-0.5,0.5,"AbsTol", 1e-5, "RelTol", 1e-3);
% Display the result
disp('Torsional stiffness');
disp(torsion);
% Integrand
function f = f_integrand(x,y,results,kappa,delta,nu_l,nu_r)
[dudx, dudy] = evaluateGradient(results, x, y);
dudx = reshape(dudx, size(x));
dudy = reshape(dudy, size(y));
young = kappa * (0.5-y).^delta + 1 -(0.5-y).^delta;
nu = nu_l * (0.5-y).^delta + nu_r * (1 -(0.5-y).^delta);
mu = young ./ (2 * (1+nu));
f = mu .* ((dudx - y).^2+(dudy + x).^2);
end
% Top boundary
function bc = bcfuncxp(location,state,kappa,delta,nu_l,nu_r)
x = location.x;
y = location.y;
young = kappa * (0.5-y).^delta + 1 -(0.5-y).^delta;
nu = nu_l * (0.5-y).^delta + nu_r * (1 -(0.5-y).^delta);
mu = young ./ (2 * (1+nu)); 
bc = mu.*x;
scatter(x,y,"filled","red"); 
hold on 
end
% Bottom boundary
function bc = bcfuncxm(location,state,kappa,delta,nu_l,nu_r)
x = location.x;
y = location.y;
young = kappa * (0.5-y).^delta + 1 -(0.5-y).^delta;
nu = nu_l * (0.5-y).^delta + nu_r * (1 -(0.5-y).^delta);
mu = young ./ (2 * (1+nu)); 
bc = -mu .* x;
scatter(x,y,"filled","red");  
hold on 
end
% Right boundary
function bc = bcfuncyp(location,state,kappa,delta,nu_l,nu_r)
x = location.x;
y = location.y;
young = kappa * (0.5-y).^delta + 1 -(0.5-y).^delta;
nu = nu_l * (0.5-y).^delta + nu_r * (1 -(0.5-y).^delta);
mu = young ./ (2 * (1+nu));
bc = mu.*y;
scatter(x,y,"filled","red"); 
hold on 
end
% Left boundary
function bc = bcfuncym(location,state,kappa,delta,nu_l,nu_r)
x = location.x;
y = location.y;
young = kappa * (0.5-y).^delta + 1 -(0.5-y).^delta;
nu = nu_l * (0.5-y).^delta + nu_r * (1 -(0.5-y).^delta);
mu = young ./ (2 * (1+nu));
bc = -mu.*y;
scatter(x,y,"filled","red");  
hold on 
end
% c-coefficient
function c = ccoeffunction(location,state,kappa,delta,nu_l,nu_r)
x = location.x;
y = location.y;
young = kappa * (0.5-y).^delta + 1 -(0.5-y).^delta;
nu = nu_l * (0.5-y).^delta + nu_r * (1 -(0.5-y).^delta);
c = young ./ (2 * (1+nu));
end
% f-function
function f = fcoeffunction(location,state,kappa,delta,nu_l,nu_r)
x = location.x;
y = location.y;
young = kappa * (0.5-y).^delta + 1 -(0.5-y).^delta;
nu = nu_l * (0.5-y).^delta + nu_r * (1 -(0.5-y).^delta);
youngpr = delta * (1-kappa)*(0.5-y).^(delta-1);
nupr = delta * (nu_r-nu_l)*(0.5-y).^(delta-1);
f = x .* (2*(1+nu) .* youngpr- 2 * young .* nupr) ./(2*(1+nu)).^2;
end
\end{lstlisting}

\subsubsection{Dual formulation}
\begin{lstlisting}
clear;
clc;
% Create a model
model = createpde();
% Include a rectangle of width ar in the model and plot the geometry
ar = 1;
R1 = [3 
      4 
      -ar/2 
      ar/2 
      ar/2 
      -ar/2 
      0.5 
      0.5 
      -0.5 
      -0.5];
g = decsg(R1);
geometryFromEdges(model,g);
figure(1)
pdegplot(model,"EdgeLabels","on") 
xlim([-ar/2-0.1 ar/2+0.1]) 
ylim([-0.6 0.6])
% Generate a mesh with a maximum edge length of 0.01. Plot the mesh.
generateMesh(model,"Hmax",0.01, "GeometricOrder", "quadratic"); 
figure(2) 
pdemesh(model)
% Specify the boundary condition 
applyBoundaryCondition(model, 'dirichlet', 'Edge', 1:4, 'u', 0);
% Specify the PDE coefficients
kappa=0.5;
delta=0.5;
nu_l=0.1;
nu_r=0.4;
ccoeff = @(location,state)ccoeffunction(location,state,kappa,delta,nu_l,nu_r);
specifyCoefficients(model,"m",0,"d",0,"c",ccoeff,"a",0,"f",2);
% Solve the equation and plot the solution.
model.SolverOptions.AbsoluteTolerance = 1e-5;  % Set absolute tolerance
model.SolverOptions.RelativeTolerance = 1e-3;  % Set relative tolerance
results = solvepde(model);
u = results.NodalSolution;
figure(3)
pdeplot(model,"XYData",u)
title("Numerical Solution");
xlabel("x")
ylabel("y")
% Retrieve the mesh data
p = model.Mesh.Nodes;
t = model.Mesh.Elements;
% Compute the torsional stiffness
func = @(x,y) f_integrand(x,y,results,kappa,delta,nu_l,nu_r);
torsion = integral2(func,-ar/2,ar/2,-0.5,0.5,"AbsTol", 1e-5, "RelTol", 1e-3);
% Display the result
disp('Torsional stiffness');
disp(torsion);
% Integrand
function f = f_integrand(x,y,results,kappa,delta,nu_l,nu_r)
[dudx, dudy] = evaluateGradient(results, x, y);
dudx = reshape(dudx, size(x));
dudy = reshape(dudy, size(y));
young = kappa * (0.5-y).^delta + 1 -(0.5-y).^delta;
nu = nu_l * (0.5-y).^delta + nu_r * (1 -(0.5-y).^delta);
mu = young ./ (2 * (1+nu));
term1 = 2 * (dudx .* x + dudy .* y);
term2 = (1 ./ mu) .* (dudx.^2 + dudy.^2); 
f = - term1 - term2;
end
% c-coefficient
function c = ccoeffunction(location,state,kappa,delta,nu_l,nu_r)
y = location.y;
young = kappa * (0.5-y).^delta + 1 -(0.5-y).^delta;
nu = nu_l * (0.5-y).^delta + nu_r * (1 -(0.5-y).^delta);
mu = young ./ (2 * (1+nu));
c = 1 ./ mu;
end
\end{lstlisting}

\subsection{Plane strain cross-sectional problem}\label{A2}

\subsubsection{Primal formulation}
\begin{lstlisting}
clear;
clc;
% Create a model
model = createpde(2);
% Create and plot geometry of a rectangle
ar = 1;
R1 = [3 
      4 
      -ar/2 
      ar/2 
      ar/2 
      -ar/2 
      0.5 
      0.5 
      -0.5 
      -0.5];
g = decsg(R1);
geometryFromEdges(model,g);
figure(1)
pdegplot(model,"EdgeLabels","on") 
xlim([-ar/2-0.1 ar/2+0.1]) 
ylim([-0.6 0.6])
% Generate a mesh with a maximum edge length of 0.01. Plot the mesh.
generateMesh(model,"Hmax",0.01,"GeometricOrder","quadratic"); 
figure(2) 
pdemesh(model)
% Define inputs for elastic moduli
kappa = 0.5;     % Young's moduli ratio 
nu_l = 0.1;     % Poisson's ratio of the left material
nu_r = 0.4;     % Poisson's ratio of the right material
delta = 2;       % Gradient index
ccoeff = @(location,state)ccoeffunction(location,state,kappa,delta,nu_l,nu_r);
fcoeff = @(location,state)fcoeffunction(location,state,kappa,delta,nu_l,nu_r);
specifyCoefficients(model,"m",0,"d",0,"c",ccoeff,"a",0,"f",fcoeff);
% Specify the boundary condition 
applyBoundaryCondition(model,"neumann", "Edge",1, "g",...
    @(location,state)traction_1(location,state,kappa,delta,nu_l,nu_r)); 
applyBoundaryCondition(model,"neumann", "Edge",2, "g",...
    @(location,state)traction_2(location,state,kappa,delta,nu_l,nu_r));
applyBoundaryCondition(model,"neumann", "Edge",3, "g",...
    @(location,state)traction_3(location,state,kappa,delta,nu_l,nu_r)); 
applyBoundaryCondition(model,"neumann", "Edge",4, "g",...
    @(location,state)traction_4(location,state,kappa,delta,nu_l,nu_r));
% Solve the equation and plot the solution.
results = solvepde(model);
u = results.NodalSolution(:,1);
dudx = results.XGradients(:,1); dvdx = results.XGradients(:,2);
dudy = results.YGradients(:,1); dvdy = results.YGradients(:,2);
xx_strain = dudx;
yy_strain = dvdy;
xy_strain = (dvdx+dudy)./2;
% Plot strain magnitude
figure
pdeplot(results.Mesh,XYData=xx_strain,ColorMap="jet")
axis equal 
title("Normal Strain Along x-Direction")
% Compute the integral
func = @(x,y) f_integrand(x,y,results,kappa,delta,nu_l,nu_r);
e_perp = integral2(func,-ar/2,ar/2,-0.5,0.5,'AbsTol', 1e-7, 'RelTol', 1e-4);
% Display the result
disp('Normalized stiffness e_perp');
disp(e_perp)
% Integrand
function f = f_integrand(x,y,results,kappa,delta,nu_l,nu_r)
[gradx, grady] = evaluateGradient(results, x, y,[1,2]);
eps_xx = reshape(gradx(:,1), size(x));
eps_yy = reshape(grady(:,2), size(y));
eps_xy = (reshape(gradx(:,2), size(x))+reshape(grady(:,1), size(y)))./2;
young = kappa * (0.5-y).^delta + 1 -(0.5-y).^delta;
nu = nu_l * (0.5-y).^delta + nu_r * (1 -(0.5-y).^delta);
mu = young ./ (2 * (1+nu));
lambda = young .* nu ./ ((1+nu).*(1-2*nu));
f = lambda .* (eps_xx + eps_yy + 2 .* nu).^2 +...
    2 * mu .* ( eps_xx + nu).^2 +...
    2 * mu .* ( eps_yy + nu).^2 +...
    4 * mu .* (eps_xy).^2;
end
% Plane strain elastic matrix C
function C = ccoeffunction(location,~,kappa,delta,nu_l,nu_r)
y = location.y;
n1 = 10;
nr = numel(y);
C = zeros(n1,nr);
young = kappa * (0.5-y).^delta + 1 -(0.5-y).^delta;
nu = nu_l * (0.5-y).^delta + nu_r * (1 -(0.5-y).^delta);
mu = young ./ (2 * (1+nu));
lambda = young .* nu ./ ((1+nu).*(1-2*nu));
% C = [2*mu+lambda; 0; mu; 0; mu; lambda; 0; mu; 0; 2*mu+lambda];
% see: https://www.mathworks.com/help/pde/ug/c-coefficient-for-systems-for-specifycoefficients.html
% case 2N(2N+1)/2-Element Column Vector c, 2-D Systems, where N=2
C(1,:) = 2*mu+lambda;
C(2,:) = zeros(1,nr);
C(3,:) = mu;
C(4,:) = C(2,:);
C(5,:) = C(3,:);
C(6,:) = lambda;
C(7,:) = C(2,:);
C(8,:) = C(3,:);
C(9,:) = C(2,:);
C(10,:)= C(1,:);
end
% Define Body Force (f-vector)
function f = fcoeffunction(location,~,kappa,delta,nu_l,nu_r)
y = location.y;
nr = numel(y);
f = zeros(2,nr); % Allocate f
young = kappa * (0.5-y).^delta + 1 -(0.5-y).^delta;
nu = nu_l * (0.5-y).^delta + nu_r * (1 -(0.5-y).^delta);
youngpr = delta * (1-kappa) * (0.5-y).^(delta-1);
nupr = delta * (nu_r-nu_l) * (0.5-y).^(delta-1);
lambdapr = ((1+nu).*(1-2*nu).*(youngpr.*nu+young.*nupr)+young.*nu.*(1+4*nu)...
    .*nupr) ./ ((1+nu).^2 .* (1-2*nu).^2);
f(1,:)= zeros(1,nr);
f(2,:)= lambdapr;
end
% Apply Surface Tractions (Neumann BCs)
function T = traction_1(location,~,kappa,delta,nu_l,nu_r)
y = location.y;
nr = numel(y);
young = kappa * (0.5-y).^delta + 1 -(0.5-y).^delta;
nu = nu_l * (0.5-y).^delta + nu_r * (1 -(0.5-y).^delta);
lambda = young .* nu ./ ((1+nu).*(1-2*nu));
T(1,:) =zeros(1,nr);
T(2,:) = -lambda;
end % Traction on the top edge
function T = traction_3(location,~,kappa,delta,nu_l,nu_r)
y = location.y;
nr = numel(y);
young = kappa * (0.5-y).^delta + 1 -(0.5-y).^delta;
nu = nu_l * (0.5-y).^delta + nu_r * (1 -(0.5-y).^delta);
lambda = young .* nu ./ ((1+nu).*(1-2*nu)); 
T(1,:) =zeros(1,nr);
T(2,:) = lambda;
end % Traction on the bottom edge
function T = traction_2(location,~,kappa,delta,nu_l,nu_r)
y = location.y;
nr = numel(y);
young = kappa * (0.5-y).^delta + 1 -(0.5-y).^delta;
nu = nu_l * (0.5-y).^delta + nu_r * (1 -(0.5-y).^delta);
lambda = young .* nu ./ ((1+nu).*(1-2*nu)); 
T(1,:) = -lambda;
T(2,:) = zeros(1,nr);
end % Traction on the right edge
function T = traction_4(location,~,kappa,delta,nu_l,nu_r)
y = location.y;
nr = numel(y);
young = kappa * (0.5-y).^delta + 1 -(0.5-y).^delta;
nu = nu_l * (0.5-y).^delta + nu_r * (1 -(0.5-y).^delta);
lambda = young .* nu ./ ((1+nu).*(1-2*nu)); 
T(1,:) = lambda;
T(2,:) = zeros(1,nr);
end % Traction on the left edge
\end{lstlisting}

\subsubsection{Dual formulation}
\begin{lstlisting}
clear;
clc;
% Create a model
model = createpde(2);
% Create and plot geometry of a rectangle
ar = 1;
R1 = [3 
      4 
      -ar/2 
      ar/2 
      ar/2 
      -ar/2 
      0.5 
      0.5 
      -0.5 
      -0.5];
g = decsg(R1);
geometryFromEdges(model,g);
figure(1)
pdegplot(model,"EdgeLabels","on") 
xlim([-ar/2-0.1 ar/2+0.1]) 
ylim([-0.6 0.6])
% Generate a mesh with a maximum edge length of 0.01. Plot the mesh.
generateMesh(model,"Hmax",0.005,"GeometricOrder","quadratic"); 
figure(2) 
pdemesh(model)
% Define inputs for elastic moduli
kappa = 0.5;     % Young's moduli ratio 
nu_l = 0.1;     % Poisson's ratio of the left material
nu_r = 0.4;     % Poisson's ratio of the right material
delta = 2;      % Gradient index
theta    = 1e6;    % penalty
ccoeff = @(location,state)ccoeffunction(location,state,kappa,delta,nu_l,nu_r,theta);
fcoeff = @(location,state)fcoeffunction(location,state,delta,nu_l,nu_r);
specifyCoefficients(model,"m",0,"d",0,"c",ccoeff,"a",0,"f",fcoeff);
% Specify the boundary condition 
applyBoundaryCondition(model, 'dirichlet', 'Edge', 1:4, 'u', [0,0]);
% Solve the equation and plot the solution.
results = solvepde(model);
u = results.NodalSolution;
dudx = results.XGradients(:,1); dvdx = results.XGradients(:,2);
dudy = results.YGradients(:,1); dvdy = results.YGradients(:,2);
% Plot data magnitude
figure
pdeplot(results.Mesh,XYData=u,ColorMap="jet")
axis equal 
% Compute the integral
func = @(x,y) f_integrand(x,y,results,kappa,delta,nu_l,nu_r,theta);
e_perp = integral2(func,-ar/2,ar/2,-0.5,0.5,'AbsTol',1e-8,'RelTol',1e-5);
% Display the result
disp('Normalized stiffness e_perp');
disp(e_perp)
% Integrand 
function f = f_integrand(x,y,results,kappa,delta,nu_l,nu_r,theta)
[gradx, grady] = evaluateGradient(results, x, y,[1,2]);
dudx = reshape(gradx(:,1), size(x));
dvdy = reshape(grady(:,2), size(y));
dudy = reshape(gradx(:,2), size(x));
dvdx = reshape(grady(:,1), size(y));
young = kappa * (0.5-y).^delta + 1 -(0.5-y).^delta;
nu = nu_l * (0.5-y).^delta + nu_r * (1 -(0.5-y).^delta);
mu = young ./ (2 * (1+nu));
f = 2 * nu .* (dudx + dvdy) ...
    -(1./(2*mu)).*( dudx.*dudx + dvdy.*dvdy + 0.5 .* (dudy + dvdx).^2) ...
    +(nu./(2*mu)).* (dudx + dvdy).^2 ...
    - 2 * theta .* (dudy - dvdx).^2;
end
% Plane strain elastic matrix C (c-matrix)
function C = ccoeffunction(location,~,kappa,delta,nu_l,nu_r,theta)
y = location.y;
n1 = 10;
nr = length(y);
C = zeros(n1,nr);
young = kappa * (0.5-y).^delta + 1 -(0.5-y).^delta;
nu = nu_l * (0.5-y).^delta + nu_r * (1 -(0.5-y).^delta);
mu = young ./ (2 * (1+nu));
% see: https://www.mathworks.com/help/pde/ug/c-coefficient-for-systems-for-specifycoefficients.html
% case 2N(N+1)/2-Element Column Vector c, 2-D Systems, where N=2
C(1,:) = (1-nu)./ (2*mu);
C(2,:) = zeros(1,nr);
C(3,:) = 1 ./ (4*mu) + 2*theta;
C(4,:) = C(2,:);
C(5,:) = 1 ./ (4*mu) - 2*theta;
C(6,:) = -nu ./ (2*mu);
C(7,:) = C(2,:);
C(8,:) = C(3,:);
C(9,:) = C(2,:);
C(10,:)= C(1,:);
end
% Define Body Force (f-vector)
function f = fcoeffunction(location,~,delta,nu_l,nu_r)
N = 2; % Number of equations
y = location.y;
nr = length(y);
f = zeros(N,nr); % Allocate f
nupr = delta * (nu_r-nu_l) * (0.5-y).^(delta-1);
f(1,:)= zeros(1,nr);
f(2,:)= -nupr;
end
\end{lstlisting}
\end{appendices}

\section*{Declaration of generative AI and AI-assisted technologies in the writing process}

During the preparation of this work the authors used Google Gemini in order to improve language and readability. After using this tool/service, the authors reviewed and edited the content as needed and take full responsibility for the content of the publication.


\end{document}